\documentclass[twocolumn,pra,aps,superscriptaddress,floatfix,showpacs]{revtex4-1}

\usepackage{amsthm}
\usepackage{amsmath}
\usepackage{graphicx}
\usepackage{amssymb}
\usepackage{bm}
\usepackage{latexsym}

\begin{document}

\date{\today}
\author{Jesse Mumford}
\affiliation{Department of Physics and Astronomy, McMaster University,
  1280 Main St.\ W., Hamilton, ON, L8S 4M1, Canada}
\author{Jonas Larson}
\affiliation{Department of Physics, Stockholm University, AlbaNova
  University Center, Se-106 91 Stockholm, Sweden}
\affiliation{Institut f\"{u}r Theoretische Physik, Universit\"{a}t zu
  K\"{o}ln, K\"{o}ln, De-50937, Germany}
\author{D. H. J. O'Dell}
\affiliation{Department of Physics and Astronomy, McMaster University, 1280 Main St.\ W., Hamilton, ON, L8S 4M1, Canada}

\title{Impurity in a bosonic Josephson junction: swallowtail loops, chaos, self-trapping and the
  poor man's Dicke model}

\begin{abstract}
We study a model describing $N$ identical bosonic atoms trapped in a double-well potential together with a single impurity atom, comparing and contrasting it throughout with the Dicke model. As the boson-impurity coupling strength is varied, there is a
symmetry-breaking pitchfork bifurcation which is analogous to the quantum phase transition occurring in the Dicke model.  Through stability analysis around the bifurcation point, we show that the critical value of the coupling
strength has the same dependence on the parameters as the critical coupling value in the Dicke model.  We also show that, like
the Dicke model, the mean-field dynamics go from being regular to chaotic above the bifurcation and macroscopic excitations of the bosons are observed.  Overall, the boson-impurity system behaves like a poor man's version of the Dicke model.   
\end{abstract}

\pacs{03.75.Lm, 05.45.Mt, 03.75.Gg, 03.65.Ta}

\maketitle

\section{Introduction}
\label{sec:introduction}
The system comprising of a single quantum particle tunneling in the presence of a many-particle environment is of fundamental interest in the study of decoherence and is relevant to certain well-known models such as the spin-boson model and the Kondo problem \cite{caldeira81,leggett87}. 
In this paper we consider a trapped ultracold atom version: a single distinguishable `impurity' atom and $N$ indistinguishable bosons all trapped together in a double-well potential. Within the single-band two-site Bose-Hubbard model both the impurity and the bosons become two level systems, i.e.\ pseudo-spins. This model has previously been studied by Rinck and Bruder \cite{rinck11}, by us \cite{mulansky11}, and by Lu and co-workers \cite{lu12}. Closely related but distinct models that have been studied recently include the cases of an impurity atom trapped in a double-well and coupled to a uniform Bose-Einstein condensate (BEC) \cite{cirone09}, an atomic quantum dot acting as a coherent single-atom or photon shuttle between two BECs \cite{bausmerth07} or two optical resonator modes~\cite{jonas0}, respectively, and of two impurities immersed in a BEC \cite{mcendoo13}. Also related are studies of double-wells containing BECs of two different species, a system suited to investigating the quantum aspects of phase separation \cite{ng03,xu08,zin11}. Away from the immediate arena of cold atoms, essentially the same Hamiltonian as we shall use here occurs in the Mermin central-spin model (also known as the spin star model) which can be pictured as a central distinguished spin  coupled equally to $N$ surrounding spins located on the points of a star \cite{breuer04,alhassanieh06,garmon11}.

Various elements of our proposed system are already well established in the laboratory, although combining them may of course prove challenging.  For example, tunnel coupled atomic BECs (bosonic Josephson junctions) have been realized in a variety of different ways including the case where the double-well might be an actual external potential \cite{cataliotti01,shin04,wang05,albiez05,gati06,schumm05,levy07,maussang10,baumgartner10,leblanc2011}, or be formed from two hyperfine states whose coupling is controlled by microwave/radio frequency fields (internal Josephson effect) \cite{gross10,zibold10}. Although binary mixtures of BECs in the same trap were first made in the early days of atomic BEC \cite{hall98}, placing just one or a controllable number of atoms in a trap is harder but can now be done \cite{serwane11}.  One set-up which is quite close to the one we have in mind here was achieved in an experiment where an optical lattice containing  a Bose-Fermi mixture was suddenly ramped up to a large depth \cite{will2011}. This resulted in an array of traps each containing either one or zero fermions together with a small number of coherent bosons. The depth of the lattice effectively shut off tunnel coupling between the wells in that experiment but ramping to smaller lattice depths would leave tunneling switched on. We also note in this context that optical lattices are versatile enough that they can be manipulated to produce a lattice of double-wells~\cite{sebby06}. 

In our previous paper \cite{mulansky11} we studied the symmetry breaking bifurcation that occurs in the ground state above a critical value of the boson-impurity interaction strength. The symmetry that is broken is a $Z_{2}$ parity symmetry whose physical order parameter is the expectation value of the difference in the number of bosons between the left and right wells (or the corresponding quantity for the impurity) which spontaneously develops a non-zero value at the bifurcation.  From the energetic point of view, above the critical interaction strength it becomes preferable for the impurity to localize in one well and for the bosons to favor the other (assuming a repulsive boson-impurity interaction) leading to a number imbalance.  Closely related symmetry breaking bifurcations have been studied experimentally in purely bosonic Josephson junctions (no impurity) \cite{zibold10}, in spin-orbit coupled BECs~\cite{spielman}, and in BECs in cavities \cite{baumann11}. In the case of \cite{zibold10} the bifurcation arises from the nonlinearity due to boson-boson interactions  \cite{botet83,milburn97,smerzi97,karkuszewski01,kanamoto06,shchesnovich09,oles10,buonsante11,juliadiaz10} and is thought to become a full blown quantum phase transition (QPT) in the limit that $N \rightarrow \infty$ \cite{botet83,shchesnovich09,oles10,buonsante11,juliadiaz10}. 
When the interactions are attractive this bifurcation occurs in the ground state, spontaneously breaking the symmetry so that the bosons clump together in either the left or right well. When the interactions are repulsive the bifurcation breaks the symmetry of excited states and manifests itself physically as the transition from Josephson oscillations to macroscopic self-trapping, i.e.\ a dynamical phase transition \cite{juliadiaz10}.
 In contrast to the purely bosonic Josephson junction, in our system it is the nonlinearity due to the boson-impurity interaction that leads to symmetry breaking and this can occur in the ground state for either repulsive or attractive interactions. Self-trapping due to the boson-impurity interaction can also occur as we shall see.

For a perfectly balanced double-well the number difference symmetry is only broken in the mean-field theory: in the fully quantum treatment the many-body wave function in Fock space (number difference space) develops non-gaussian number fluctuations and eventually separates into two macroscopically distinguishable pieces, i.e.\ a Schr\"{o}dinger cat state. This state is notoriously delicate and tiny external perturbations not included in the Hamiltonian are liable to break the symmetry by effectively introducing a tilt between the two wells. This collapses the cat state and thereby restores the validity of the mean-field result. Another difference between the full quantum theory and the mean-field theory is that the latter is nonlinear and this is the origin of the bifurcation which takes the form of a 3-pronged pitchfork when the number difference is plotted as a function of the boson-impurity interaction strength and appears as a swallowtail loop \cite{mulansky11} (see Figure \ref{fig:glNoU} below) when the energy is plotted versus an externally imposed tilt (which plays a role analogous to quasimomentum \cite{krahn09}).  These iconic loop structures also occur in many other bifurcating systems including bosonic Josephson junctions \cite{karkuszewski01}, the band structure of BECs in optical lattices \cite{wu01,diakonov02,mueller02} including at the Dirac point for a honeycomb lattice \cite{chen11}, the band structure of non-interacting atoms in cavity-QED \cite{prasanna11}, and the equivalent of band structure for BECs in toroidal traps \cite{mason09,baharian13}. Their presence has also been inferred experimentally due to a sudden break down in adiabaticity during a parametric sweep of the tilt between wells in a bosonic Josephson junction  \cite{chen11b}.

In this paper we shall show that at the same time the bifurcation appears the mean-field dynamics goes from regular to chaotic. This feature does not occur in the ordinary bosonic Josephson junction (when treated within the standard two-mode model). Indeed, by adding an additional degree of freedom like a spin chaos has been predicted to appear in both mean-field BECs~\cite{ghose01} and bosonic Josephson junctions~\cite{horsdal}. Alternatively, chaos may also occur by removing energy conservation \cite{weiss08,lee01,abdullaev00}. Chaos is a classical phenomenon that is usually defined as exponential sensitivity to initial conditions, that is, two arbitrarily close points in phase
space will diverge exponentially over time.  In quantum mechanics, precise trajectories do not exist and positions in phase space cannot be defined than better to an area of size $\approx \hbar$ precluding the possibility of exponential sensitivity. Nevertheless, quantum systems whose classical limit is chaotic do display tell-tale behaviour such as level repulsion leading to the idea of quantum chaology \cite{berry87,berry88}. Here we demonstrate chaos via Poincar\'{e} plots giving stroboscopic sections through classical (mean-field) phase space and also by monitoring the statistics of the quantum energy levels.

Similar regular-to-chaotic behavior as we observe has recently been predicted in the celebrated Dicke model \cite{altland12}. Indeed, in this paper we make the claim that our model is a poor man's version of the Dicke model, behaving identically if one is close to the bifurcation. The original Dicke model described $N$ two-level atoms coupled to a single mode of the electromagnetic field and undergoes a quantum phase transition to a superradiant phase corresponding to the collective emission of photons at a critical value of the atom-light coupling strength \cite{dicke54}.  Alternatively stated, the Dicke model consists of $N$ spins coupled to a harmonic oscillator. The physical basis of our claim of the equivalence of the two models is that very near the critical point the harmonic oscillator is barely excited and can be truncated to just two states: its ground and first excited state, and therefore behaves like the two-state impurity atom in our model.  The fact that at a phase transition quantum fluctuations become important but that this point also coincides with the onset of chaos, which is a classical phenomenon, suggests intriguing connections between the quantum and classical worlds \cite{ghose05}.

The layout of this paper is as follows: We introduce the boson-impurity and the Dicke
hamiltonians in Section \ref{sec:Model}; Sec.~\ref{sec:sa} consists of an analysis of the stationary mean-field problem including loops and the stability of the solutions; In Sec.~\ref{sec:mfc} we show the emergence of classical chaos which is triggered by the bifurcation and hence the presence of the impurity; the following Sec.~\ref{sec:selft} demonstrates self-trapping, and in Sec.~\ref{sec:ls} we analyze the nearest neighbour statistics of the quantum energy levels which further illustrates that chaos is to be expected in the classical limit.  We have also provided two appendices: Appendix \ref{sec:appendixparity} explains how we ensure that the eigenstates produced by numerical diagonalization have well defined parity and Appendix \ref{sec:app} contains Poincar\'{e} sections through classical phase space that illustrate how the dynamics changes as we sample different energies.

\section{Model}
\label{sec:Model}
Within the  two-site single band Bose-Hubbard model, i.e.\ the two mode model, the many-body Hamiltonian for the boson-impurity system is given by \cite{rinck11,mulansky11}
\begin{eqnarray}
\hat{H} &=&  - J \hat{B} - J^a \hat{A} +
\frac{W}{2} \Delta \hat{N} \Delta \hat{M} \nonumber \\
&& + \frac{\Delta \epsilon}{2} \Delta \hat{N} + \frac{\Delta
  \epsilon^a}{2} \Delta \hat{M} \, .
\label{eq:mbham}
\end{eqnarray}
Here, $\Delta \hat{N} \equiv \hat{b}^{\dagger}_R
\hat{b}_R - \hat{b}^{\dagger}_L \hat{b}_L$ is the number difference
operator between the two wells for the bosons, and $\hat{B} \equiv
\hat{b}^{\dagger}_L \hat{b}_R + \hat{b}^{\dagger}_R \hat{b}_L$ is the
boson hopping operator which also gives the coherence between the two wells \cite{gati07}. Likewise, $\hat{M} \equiv \hat{a}^{\dagger}_R
\hat{a}_R - \hat{a}^{\dagger}_L \hat{a}_L$ and $\hat{A} \equiv
\hat{a}^{\dagger}_L \hat{a}_R + \hat{a}^{\dagger}_R \hat{a}_L$ are
the equivalent operators for the impurity. We assume that both the boson and impurity creation/annihilation operators obey the standard bosonic commutation relations, i.e.\ $[\hat{b}_\alpha,\hat{b}_\alpha^\dagger]=[\hat{a}_\alpha,\hat{a}_\alpha^\dagger]=1$ with $\alpha=L,\,R$ and all the remaining commutators are identically zero. However, because there is only one impurity its statistics do not matter and it could be a boson or a fermion.  $W$ parameterizes the boson-impurity interaction, and $J$ and $J^a$ are the hopping amplitudes for the bosons and impurity, respectively. Using similar notation, $\Delta \epsilon$ and $\Delta\epsilon^a$ are the respective differences between the
zero-point energies of the two wells, i.e.\ the tilt, for the bosons and the impurity. It is important to appreciate that we do not include direct boson-boson interactions in our model, assuming that they can be removed by a Feshbach resonance if necessary \cite{zibold10}. In our previous paper \cite{mulansky11} we did include them, but for many of the effects we are interested in here, especially the bifurcation in the ground state, they are a distraction that does not make a qualitative difference to the behavior. One exception to this is the particular case of attractive boson-boson interactions above a certain threshold in which case they also cause a symmetry breaking bifurcation in the ground state as discussed in the Introduction.

The Hamiltonian in Equation (\ref{eq:mbham}) can be re-expressed in a spin notation  by using the symmetric/antisymmetric (S/AS) modes instead of the left/right (L/R) modes as a basis. The S/AS modes are the eigenmodes of the single particle problem, i.e.\ in the absence of interactions. Therefore, in the limit that $W \rightarrow 0$ the ground state corresponds to all the particles in the S mode because it has lower energy. Using a simple Hadamard-rotation of the L/R creation (annihilation) operators we have,
\begin{eqnarray}
\hat{b}_L &=& \frac{1}{\sqrt{2}} \left ( \hat{b}_{S} + \hat{b}_{AS} \right ) \\
\hat{b}_R &=& \frac{1}{\sqrt{2}} \left ( \hat{b}_{S} - \hat{b}_{AS} \right ) \, 
\end{eqnarray}
and similar expressions hold for the impurity operators.
In the new basis, and for vanishing tilts $\Delta\epsilon=\Delta\epsilon^a=0$, Eq.~(\ref{eq:mbham}) takes the form 
\begin{equation}
\hat{H}_{\mathrm{S,AS}} = 2J \hat{S}_z + 2J^a \hat{S}^a_z + 2 W \hat{S}_x \hat{S}^a_x \, 
\label{eq:smallham}
\end{equation}
where the Schwinger spin-representation has been used~\cite{sakurai}, i.e.\ $\hat{S}_z \equiv ( \hat{b}^{\dagger}_{AS} \hat{b}_{AS} -
  \hat{b}^{\dagger}_{S} \hat{b}_{S} )/2 =-\hat{B}/2$ and
$\hat{S}_x \equiv  ( \hat{b}^{\dagger}_{AS} \hat{b}_{S} +
  \hat{b}^{\dagger}_{S} \hat{b}_{AS} )/2 =-\Delta \hat{N}/2$. 
Apart from the trivial $U(1)$ symmetry related to particle conservation, we note that the hamiltonian supports a $Z_2$ parity symmetry under $\hat{S}_x\rightarrow-\hat{S}_x$, $\hat{S}_y\rightarrow-\hat{S}_y$, $\hat{S}_z\rightarrow\hat{S}_z$, and equivalently for the impurity spin operators. This spin rotation preserves the $SU(2)$ angular momentum commutation relations. Note that in the original L/R-basis this symmetry is nothing but a reflection of the double-well about the origin. It follows that a non-zero tilt $\Delta\epsilon \neq 0$ or $\Delta\epsilon^a \neq 0$ breaks this symmetry.   

As claimed in the introduction, our model hamiltonian in the form of Eq.\ (\ref{eq:smallham}) shows some resemblance to the Dicke hamiltonian~\cite{dicke54}
\begin{equation}
\hat{H}_{\mathrm{D}} = \omega_B \hat{S}_z + \omega_A \hat{c}^{\dagger}\hat{c} +
2 g (\hat{c} + \hat{c}^{\dagger})\hat{S}_x \, .
\label{eq:dicke}
\end{equation}  
Here, $\hat{c}^{\dagger} (\hat{c})$ are photon, i.e.\ boson, creation (annihilation)
operators and $\hat{S}_z$ and $\hat{S}_x$ are spin operators.
Eq.~(\ref{eq:dicke}) describes a spin-$N/2$ system coupled to the position coordinate of
a harmonic oscillator.  The frequencies
$\omega_B$ and $\omega_A$ are the spin precession and harmonic
oscillator frequencies, respectively, and $g$ is the coupling
strength. This system experiences a QPT at
a certain critical value $g_c$ which will be discussed further below~\cite{dqpt}.  
By comparison, Eq.\ (\ref{eq:smallham}) can now be thought of as a system consisting of a spin-$N/2$ coupled to a spin-1/2 particle (impurity) instead of a harmonic oscillator (Dicke model). Here lies the most important distinction between the two hamiltonians: in the Dicke model the coupling is to the electromagnetic field which has infinitely many energy levels whereas there are only two
levels for the impurity. Despite this truncation of the Hilbert space of the Dicke model, the parity symmetry of the boson-impurity model has its analogue in the Dicke model with $\hat{c}\rightarrow-\hat{c}$ and $\hat{c}\dagger\rightarrow-\hat{c}\dagger$. These similarities mean that the boson-impurity system behaves as a simplified, or poor man's, Dicke model that captures the crucial behaviour near the QPT. We also note that the rotating-wave approximation has not been imposed, either in Eq.\ (\ref{eq:smallham}) or in Eq.\ (\ref{eq:dicke}).  

In order to obtain many of the results presented in this paper we numerically diagonalize the Hamiltonian given in Eq.\ (\ref{eq:mbham}). For $N$ atoms this requires diagonalizing a $(2N+2)\times (2N+2)$ matrix which is tractable for $N \sim 100$ on a small computer because of the linear scaling in $N$ of the matrix dimension and thus we can obtain numerically exact results. There are, however, some subtleties  involved which are reflected in the physics of the system: From the above discussion we see that in the absence of any tilt the eigenstates should all have well defined parity in the two-dimensional Fock space (i.e.\ number difference space) where the many-body quantum state lives. However, numerical diagonalization routines do not automatically respect this parity symmetry. The most severe test occurs in the critical region where the quantum state becomes non-gaussian and eventually evolves into a Schr\"{o}dinger cat state made of two almost separated pieces in Fock space connected only by exponentially small probability amplitudes in between. This may be viewed as arising from the appearance of an effective double well potential in Fock space: for each even parity state there is an odd parity one and their energies become almost degenerate except for an exponentially small tunnel splitting when they lie below the barrier top.   Numerical routines find it hard to handle exponentially small numbers at the same time as numbers of order unity and tend to give eigenstates of broken parity above the critical value of $W$, i.e.\ the eigenstates choose one well or the other. It is amusing to reflect on the fact that numerical errors replicate the effects of a physical environment! As we explain in Appendix \ref{sec:appendixparity}, we circumvent these problems by diagonalizing the Hamiltonian in a basis which has well defined parity so that good parity in Fock space is built in from the start.

\begin{figure}
\begin{center}
\includegraphics[width=0.7\columnwidth]{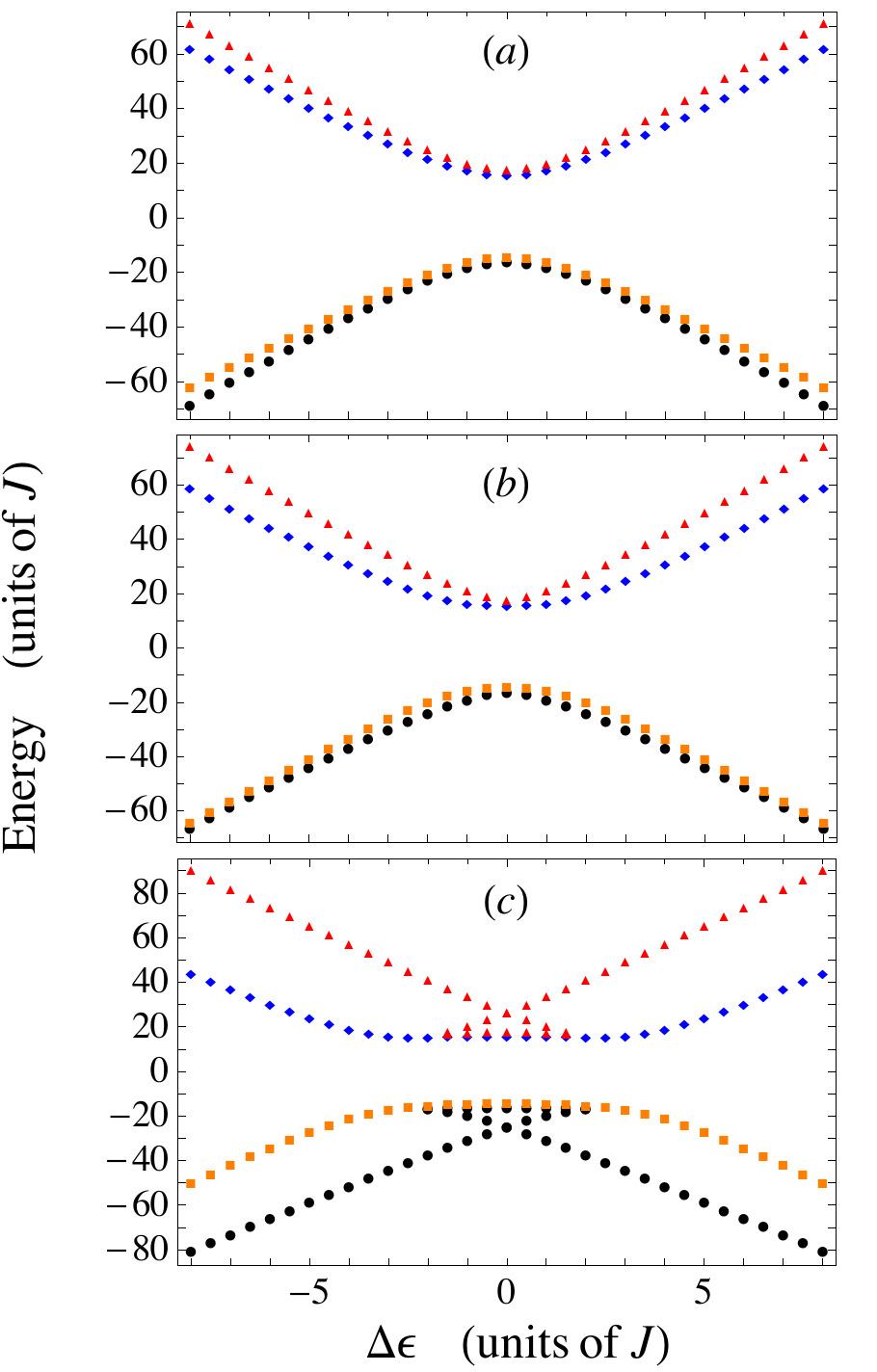}
\end{center}
\caption[Mean-field energies]{(Color online) Energies of the static solutions to the
  mean-field equations (\ref{eq:josephson1})--(\ref{eq:josephson4}) as a function of the
  tilt $\Delta \epsilon$.  Each panel has a different value
  of the boson-impurity interaction energy:  (a) $W=0.1J$, (b) $W=0.5J$, and (c)
  $W=2.5J$. The various
  solutions within each panel are characterized by their phase difference.  The four
  solutions are:
  $\alpha=\beta=0$ (black circles); $\alpha=\pi$ and $\beta=0$ (orange
  squares), $\alpha=0$ and $\beta=\pi$ (blue diamonds), and
  $\alpha=\beta=\pi$ (red triangles).  All panels have $J^{a}= J$, $\Delta \epsilon^{a}=\Delta \epsilon$ and $N=16$.  }
	\label{fig:EvTilt}
\end{figure}

\section{Mean-field analysis}
\label{sec:sa}
We perform the mean-field approximation by replacing the 
operators $\hat{a}_{L/R}$ and $\hat{b}_{L/R}$ in Eq.\ (\ref{eq:mbham})
with complex numbers
\begin{eqnarray}
\hat{a}_{L/R} &\rightarrow& a_{L/R} = \sqrt{M_{L/R}} \, e^{i \alpha_{L/R}
  (t)} \\
\hat{b}_{L/R} &\rightarrow& b_{L/R} = \sqrt{N_{L/R}} \, e^{i \beta_{L/R}
  (t)}
\end{eqnarray}
giving the mean-field hamiltonian 
\begin{eqnarray}
H_{\mathrm{MF}} &=& - J \sqrt{N^2 - 4 Z^2} \cos{\beta} - J^a \sqrt{1 - 4 Y^2}
\cos{\alpha} \nonumber \\ 
&& + 2 W Y Z + \Delta \epsilon Z + \Delta \epsilon^a Y\, . 
\label{eq:mfham}
\end{eqnarray}  
In the above expression we have introduced the classical variables giving the phase and population differences between the $L$ and $R$ sides: $\alpha \equiv \alpha_L -
\alpha_R$, $Y \equiv \Delta M/2$, $\beta \equiv \beta_L - \beta_R$,
and $Z \equiv \Delta N/2$. It should be noted that the mean-field approximation is really only being applied to the boson field. The mean-field representation for the impurity is in fact exact because the quantum state of a spin-1/2 is fully characterized by the two real numbers $\alpha$ and $Y$ which can be related to the two angles on the Bloch sphere. Here we utilize the boson coherent state ansatz to derive the semi-classical hamiltonian~(\ref{eq:mfham}), but an equally good approach would be to use spin coherent states instead~\cite{altland12}.

\begin{figure}
\begin{center}
\includegraphics[width=0.7\columnwidth]{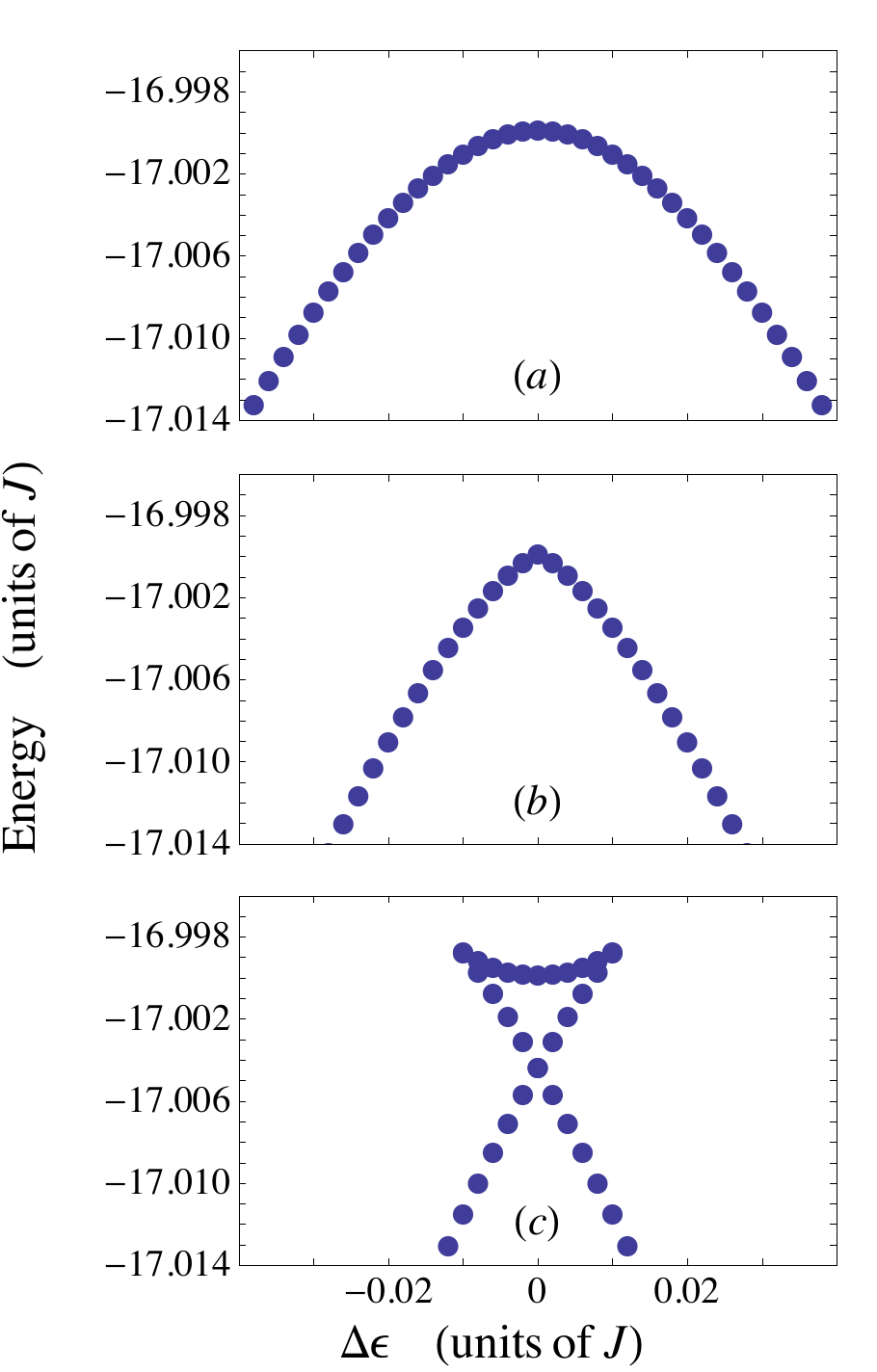}
\end{center}
\caption{(Color online) A close-up of the central region of the lowest band 
  showing the emergence of the swallowtail loop as the
  boson-impurity interaction strength $W$ is varied: (a) $W=0.475J$,
  (b) $W=0.5J$, and (c) $W=0.525J$.   The other parameters are
 $J^{a}= J$, $\Delta \epsilon^{a}=\Delta \epsilon$, and $N=16$. Of the three branches of the loop, the unstable one is the upper one, i.e.\ the curved part along the top.}
\label{fig:glNoU}
\end{figure}

The equations of motion following from Eq.\ (\ref{eq:mfham}) can be obtained from Hamilton's equations \cite{mulansky11}
\begin{eqnarray}
 \dot{\alpha} & = & \displaystyle{\frac{\Delta \epsilon^a }{\hbar} +
 2\frac{W}{\hbar} Z + \frac{4J^a }{\hbar}  \frac{Y \cos{\alpha}}{\sqrt{1-4Y^{2}}}} \label{eq:josephson1} \\ 
	 \dot{Y} &= & \displaystyle{-\frac{J^a }{\hbar} \sqrt{1-4 Y^{2}} \sin{\alpha} } \label{eq:josephson2} \\ 
	\dot{\beta} &= & \displaystyle{\frac{\Delta \epsilon}{\hbar} + 2\frac{W}{\hbar} Y + \frac{4J}{\hbar} \frac{ Z \cos{\beta} }{\sqrt{N^2-4Z^{2}}} } \label{eq:josephson3} \\ 
	  \dot{Z} &= & - \displaystyle{\frac{J}{\hbar} \sqrt{N^2-4 Z^{2}}}
          \sin{\beta} \, . 
          \label{eq:josephson4} 
	 \end{eqnarray}
Setting the left hand sides to zero gives the stationary solutions whose energies are plotted in Fig.\ \ref{fig:EvTilt} as a function of the tilt, $\Delta \epsilon$.  Ignoring the trivial solutions of $Z = \pm N/2$ and $Y = \pm 1/2$ we consider the solutions $\alpha = \{0, \pi\}$ and $\beta =
\{0, \pi\}$ to Eqs.~(\ref{eq:josephson1})--(\ref{eq:josephson4}) giving four different combinations of the phase differences.  In terms of the double pendulum analogy advanced in \cite{mulansky11}, the combination $(\alpha=\beta=0)$ corresponds to both pendula pointing straight down, $(\alpha=\pi,\beta=0)$ corresponds to the impurity pendulum pointing straight up and the boson pendulum pointing straight down and vice versa for $(\alpha=0,\beta=\pi)$. The combination $(\alpha=\pi,\beta=\pi)$  corresponds to both pendula pointing straight up. In Fig.\ \ref{fig:EvTilt} each of these solutions is plotted with a different symbol. Each panel is for a different
value of $W$ and illustrates how swallowtail loops appear when $W$ exceeds a certain critical value $W_{c}$.

Let us focus upon panel (c) of Fig.\ \ref{fig:EvTilt} which is for $W>W_{c}$ and contains two swallowtail loops, one in the lowest and one in the highest energy band. Indeed, the stationary classical states  are symmetric about the $E=0$ center line as can be seen from the figure. The area of the loops depends on $W$, but their positions along the vertical energy axis are determined by $J$ and $J^a$. At $\Delta\epsilon = \Delta\epsilon^a =0$ the top of the $\alpha = \beta = 0$ band is fixed at $E=-NJ - J^a$
even when the loop is formed. Meanwhile the next band up, the $\alpha = \pi$, $\beta = 0$ band
is fixed at  $E=-NJ + J^a$. The energy gap between the two lowest bands is therefore $2J^a$ and the same goes for energy gap between the two highest bands. The appearance of two loops is in contrast to the boson-only system where boson-boson interactions lead to a single loop, either in the highest band for the case of repulsive interactions or in the lowest band for the attractive case.

\begin{figure}
\begin{center}
\includegraphics[width=0.7\columnwidth]{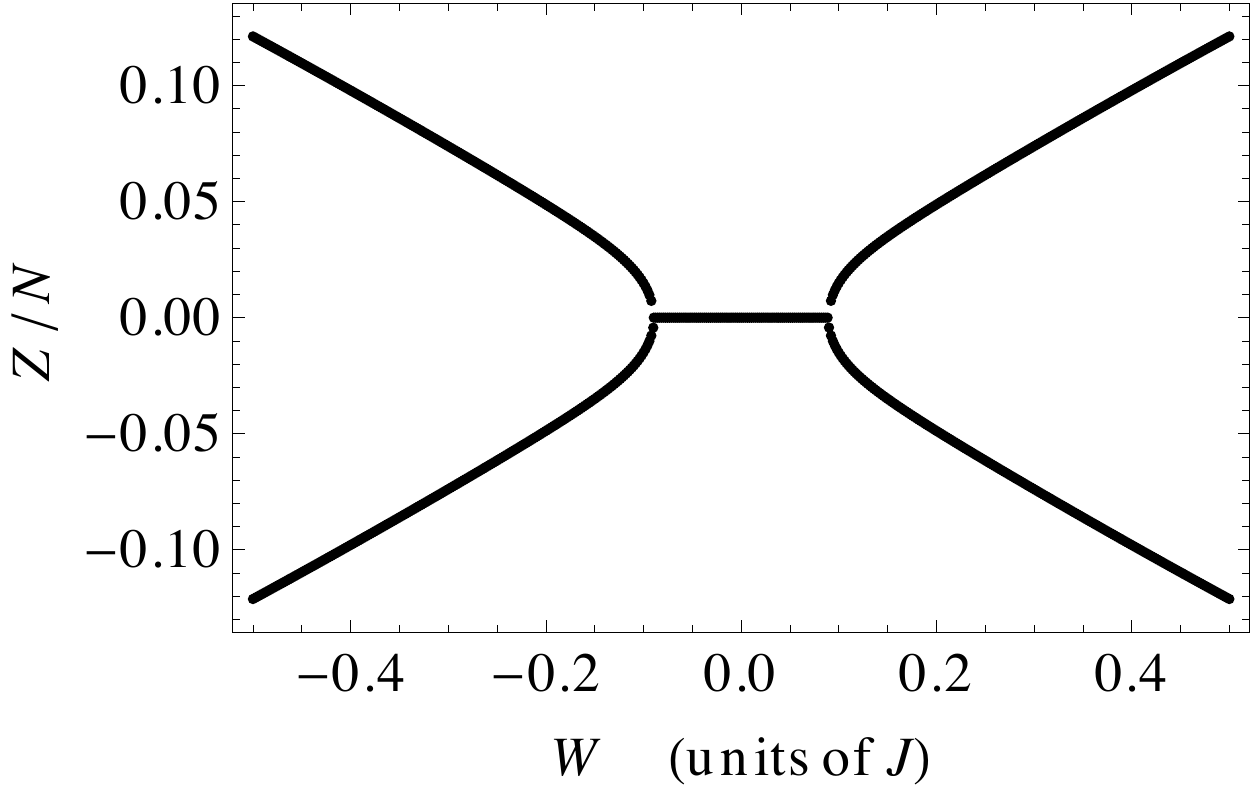}
\end{center}
	\caption{Supercritical pitchfork bifurcation in the boson number difference $Z$ between
          the right- and left-hand wells for the lowest
          band plotted as a function of the boson-impurity interaction
          strength $W$. The solid and dotted lines signify stable and
          unstable solutions, respectively.  We see that the bifurcation occurs for both repulsive and attractive boson-impurity interactions. The values of the parameters are $N=500$, $\Delta \epsilon=\Delta
          \epsilon^{a}=0$, $J^{a}=J$.}
	\label{fig:pitchforkz}

\end{figure}

To investigate the appearance of loops further we have plotted in Fig.\ \ref{fig:glNoU} an enlargement of the lowest band. In panel $(a)$ $W < W_c$ we have a smooth curve.  In panel $(b)$ $W = W_c$ a cusp forms at zero tilt heralding the
emergence of the loop. In panel $(c)$ $W > W_c$ a loop forms where the number of solutions for ($\alpha=\beta=0$)
increases from one to three for a range of tilts. Each of these three solutions are distinguished by their number differences which form a pitchfork bifurcation when plotted as a function of $W$ as illustrated in  Fig.\ \ref{fig:pitchforkz}.

In order to see analytically how $W$ causes the loop structure we perform a
stability analysis in the region where the loop first occurs.
The lower band is characterized by $\alpha = \beta = 0$ and the loop
begins to form at zero tilt.  Therefore, solving equations 
\begin{eqnarray}
 2 W Z + 4 J^a \frac{Y}{\sqrt{1-4Y^{2}}}   &=& 0 \label{eq:lowenband1} \\
	 2 W Y + 4 J \frac{Z}{\sqrt{N^2-4 Z^{2}}} &=& 0  
	\label{eq:lowenband2}
\end{eqnarray}
gives $\alpha = Y = \beta = Z = 0$ as the point at which the loop
appears.  If we define $\vec{x} = (\alpha,Y,\beta,Z)$ and linearize
$\dot{\vec{x}}$ around $\vec{x}_0 = (0,0,0,0)$ we obtain
\begin{equation}
\dot{\vec{x}}=J(\vec{x}_{0}) \vec{x}+O(\vec{x}^2)
\end{equation}
where $J(\vec{x}_{0})$ is the Jacobian matrix evaluated at
$\vec{x}_{0}$,
\begin{equation}
J(\vec{x}_{0}) =  \left[ \begin{array}{cccc} 0 & 4J^a & 0 & 2W \\ -J^a & 0
    & 0 & 0 \\ 0 & 2W & 0 & 4J/N \\ 0 & 0 & -JN & 0 \end{array}
\right] \, .
\end{equation}
The stability of the system at $\vec{x}_{0}$ depends on the
eigenvalues of $J(\vec{x}_{0})$.  Every solution at $\vec{x}_{0}$ will
be unstable if there is a positive real eigenvalue because there will
be solutions like $\vec{x}(t)=e^{\lambda t}$.  The eigenvalues are
\begin{align}
\lambda_{1,\pm} & = & -\sqrt{2 \left ( -J^2-J^{a^2} \pm
  \sqrt{\left ( J^2 - J^{a^2} \right )^2 + JJ^aW^2N} \right
)} \label{eq:lambda1} \\
\lambda_{2,\pm} & = &  \sqrt{2 \left ( -J^2-J^{a^2} \pm
  \sqrt{\left ( J^2 - J^{a^2} \right )^2 + JJ^aW^2N} \right )} \, .
\label{eq:lambda2}
\end{align}
Looking at $\lambda_{2,+}$ we find that the critical value of $W$ when
the loop emerges is
\begin{equation}
W_c = 2 \sqrt{\frac{JJ^a}{N}} \, .
\label{eq:Wcrit}
\end{equation}
Comparing this with the critical coupling strength, $g_c = \sqrt{\omega_A \omega_B/4 N}$, at which a phase transition
occurs in the Dicke model we see that they have exactly the same dependence upon the associated parameters in the two Hamiltonians given in Eqs. (\ref{eq:smallham}) and (\ref{eq:dicke}) (the factor of 4 difference
can be attributed to the definitions we use). Where
the boson-impurity system experiences a bifurcation the Dicke model
experiences a QPT in the limit that $N \rightarrow \infty$ (see also Ref.~\cite{parkins}).

The preceding analysis also provides some information on the type of
bifurcation shown in Fig.\ \ref{fig:pitchforkz}. Since the stationary
point $\vec{x}_{0}=(0,0,0,0)$ goes from being stable to unstable as
$W$ is increased through $W_c$ we have a supercritical pitchfork bifurcation.
This type of bifurcation is common for systems with symmetry
($H_{\mathrm{MF}} \rightarrow H_{\mathrm{MF}}$ for zero tilt
under $\vec{x} \rightarrow - \vec{x}$).  When $W = W_c$,
$\lambda_{1,\pm} = \lambda_{2,\pm} = 0$ and the solutions experience a
process called ``critical slowing down'' where the decay/growth time
to/from initial conditions is no longer exponential.  A pitchfork
bifurcation for $Z$ has been observed experimentally in a
bosonic Josephson junction~\cite{zibold10},  in a spin-orbit
coupled BEC~\cite{spielman}, and at the onset of a density wave instability in a BEC in an optical cavity \cite{baumann11}. In fact, this latter problem can be mapped onto the Dicke problem too \cite{baumann11,baumann10,nagy10,bhaseen12,gopal09,buchhold12}.  If a non-zero tilt is applied then the pitchfork opens up, as shown in Figs.\ 5 and 6 in \cite{mulansky11}. Considering
Fig.\ \ref{fig:glNoU} we can see that a finite tilt delays the onset of the bifurcation to a larger value of $W$ because the loop is born at $\Delta \epsilon=0$ and grows outwards.

\begin{figure}
\begin{center}
\includegraphics[width=0.6\columnwidth]{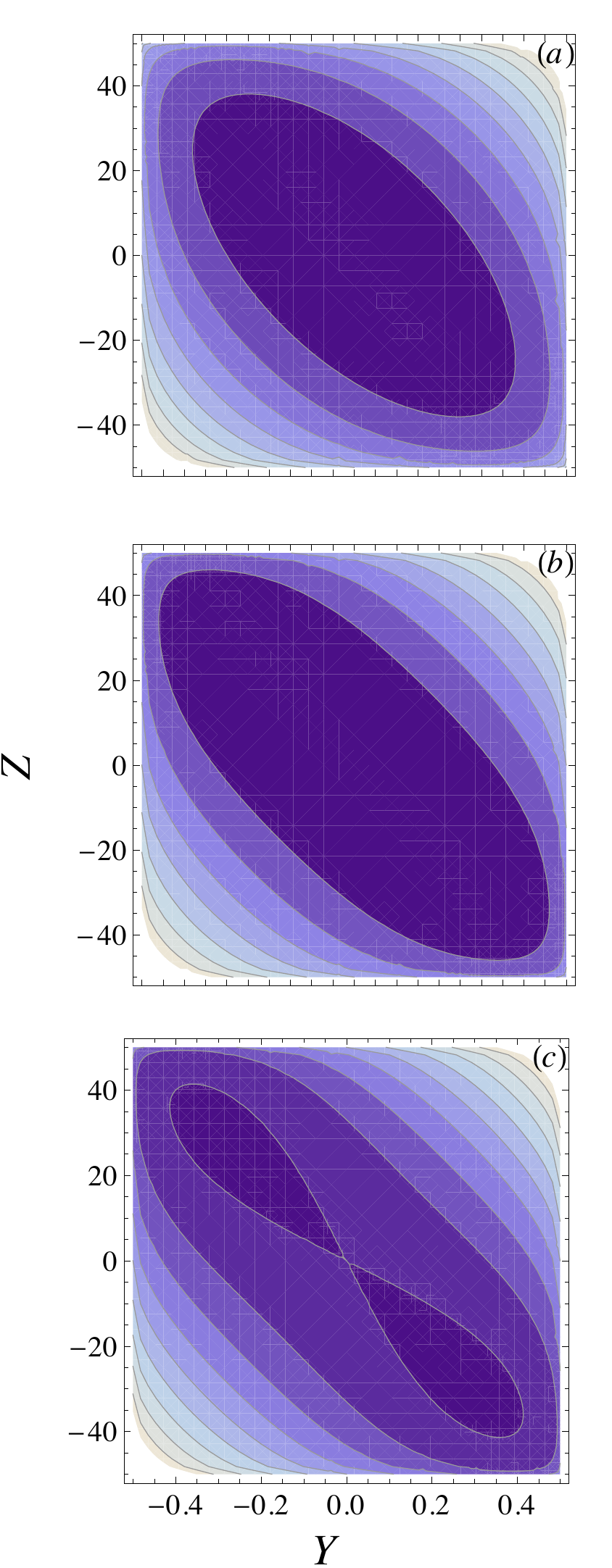}
\end{center}
	\caption{(Color online) Contour plots of the mean-field energy $H_{\mathrm{MF}}$ evaluated for the lowest band in the number difference ($YZ$)  plane. This figure shows how a double-well forms when $W>W_c$.    (a) $W = 0.75W_c$, (b) $W = W_c$ and (c) $W =
          1.25W_c$.   The other parameter values are $J^a = N J$, $N = 100$ and $\Delta \epsilon=\Delta
          \epsilon^{a}=0$.  Lighter coloured regions are higher in energy.}
	\label{fig:contourcolumn}

\end{figure}

The position in phase space of the new stable points can be found
analytically by solving Eqs.\ (\ref{eq:lowenband1}) and (\ref{eq:lowenband2}) for
$W>W_c$,
\begin{eqnarray}
\vec{x}_{\pm} = \Biggl ( &0&, \pm \frac{1}{2W}
\sqrt{\frac{N^2W^4-16J^2J^{a^2}}{N^2W^2+4J^{a^2}}}, \nonumber \\
&0&, \mp\frac{1}{2W} \sqrt{\frac{N^2W^4-16J^2J^{a^2}}{W^2+4J^2}} \Biggr
) \, .
\label{eq:YZposition}
\end{eqnarray}
As $W \rightarrow \infty$ we have $\vec{x}_{\pm} \rightarrow
(0, \pm 1/2, 0, \mp N/2)$ corresponding to the complete localization of all the particles in one well or the other, as expected for a large
interaction, attractive or repulsive, between the impurity and bosons.  Fig.\ \ref{fig:contourcolumn}  plots the mean-field energy in the $Y-Z$ plane for the lowest band
for values of $W$ close to $W_{c}$. As $W$ passes through $W_{c}$ a double well forms in \emph{Fock space} (number difference space) in accordance with the Landau model for second order phase transitions.  In Fig.\ \ref{fig:contourcolumn} this
double-well forms along the diagonal $Z = -Y$ because we have set $J^a = NJ$ which means that
the hopping energies for the impurity and bosons contribute equally to the total energy of
the system.  In general the axis of the double-well can be
at any angle in the $YZ$ plane depending on the parameters in $H_{\mathrm{MF}}$.

Defining $\chi = W/W_c$, the energies at the minima when $W>W_c$ and for zero tilt 
are 
\begin{align}
E_{|W|>W_c} = -2JN \biggl [ \frac{\sqrt{\left(\eta + \chi^2\right) \left(1+
      \eta \chi^2 \right)}}{2 \eta \chi} - 1 \biggr ] \, ,
\end{align} 
where $\eta = JN/J^a$.  When $J^a = JN$, $\eta = 1$ giving
\begin{equation}
E_{|W|>W_c} = - 2JN \Bigl [\frac{\chi}{2} \left ( 1 + \chi^{-2} \right ) - 1
\Bigl] \, .
\label{eq:groundBI}
\end{equation} 
This corresponds to the case $\omega_A = \omega N$ and $\omega_B =
\omega$ in the Dicke model where the ground state energy in the
super-radiant phase becomes \cite{nahmad-achar13}
\begin{equation}
E_{|g|>g_c} = - \frac{\omega N}{2} \Bigl [\frac{\chi^2}{2} \left ( 1 +
  \chi^{-4} \right ) - 1 \Bigr ] \, .
\label{eq:groundDicke}
\end{equation} 
We can see that Eqs.\ (\ref{eq:groundBI}) and (\ref{eq:groundDicke})
differ slightly in their dependence on $\chi$.  However, in the vicinity of the critical coupling
strengths we can set $\chi = 1 + \delta$ where $\delta \ll 1$ and
expand to leading order to give $E_{|W|>W_c}/J = E_{|g|>g_c}/\omega = - N
\delta^2$. Thus, near the critical point the ground state energies behave in
exactly the same way.     

We now shift our focus from Eq.\ (\ref{eq:mbham}) to Eq.\
(\ref{eq:smallham}) and look at the dependence of $\langle \hat{S}_z
\rangle$ in the ground state  on the driving parameter
$W$ (recall that $\hat{S}_z$ is the number difference between
the anti-symmetric and symmetric modes).  We compare our results with those obtained from the
Dicke model by Emary and Brandes \cite{emary03}. To properly compare the two systems we
scale the third term of $\hat{H}_D$ by $1/\sqrt{N}$ and the second term
of $\hat{H}_{\mathrm{S,AS}}$ by $N$, so every term in each hamiltonian is
$\mathcal{O}(N)$ (this is the only point we
perform these scalings and we return to the pre-scaling hamiltonians
starting in the next section). Thus, the scaled hamiltonian is properly defined in the thermodynamic limit and the classical pitchfork bifurcation signals the presence of a QPT. The critical coupling parameters remain constant in
the thermodynamic limit $N \rightarrow \infty$; $W_c = \sqrt{JJ^a}/2$ and
$g_c = \sqrt{\omega_A \omega_B}/2$.

\begin{figure}[t]
\begin{center}
\includegraphics[width=0.8\columnwidth]{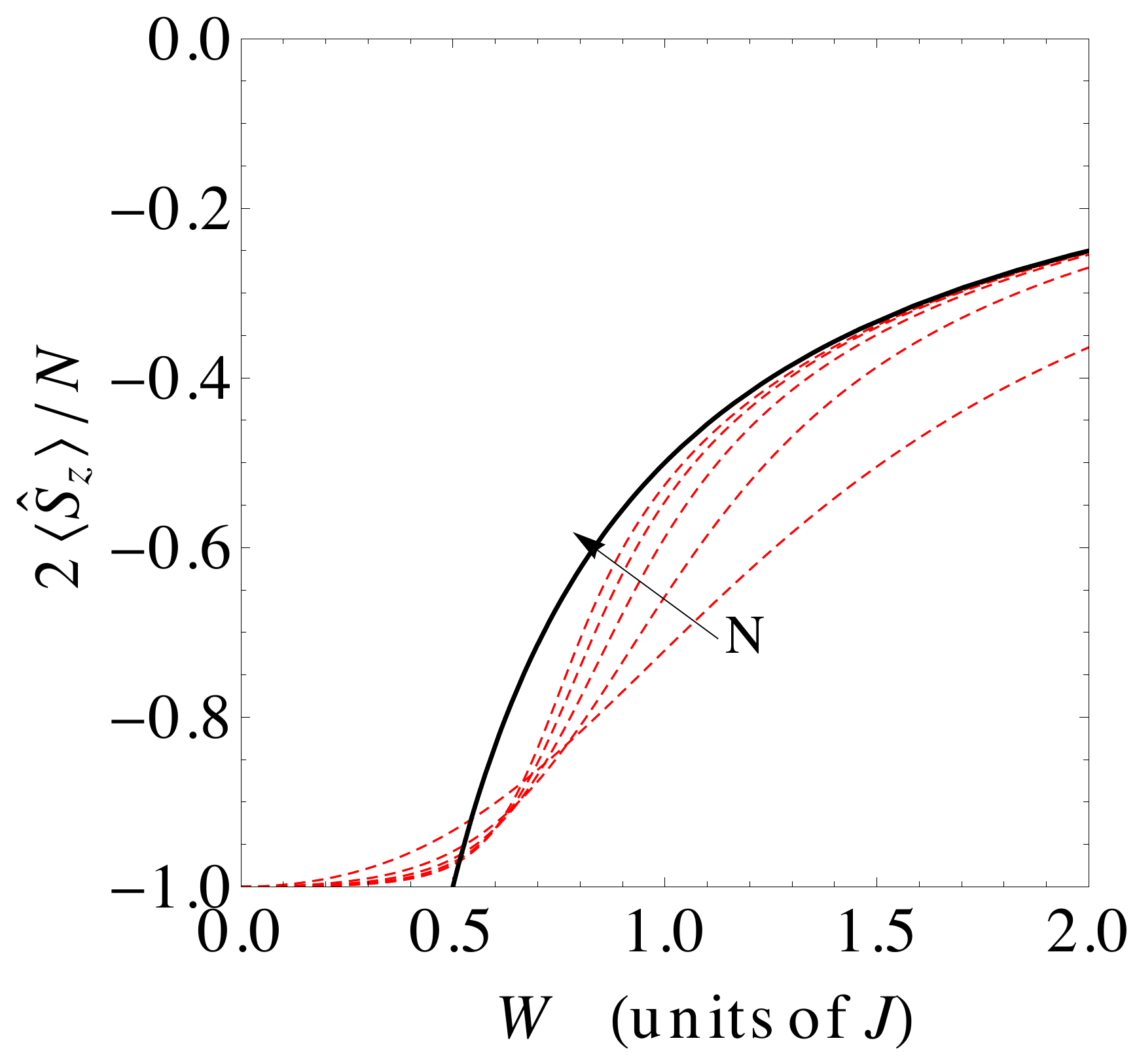}
\end{center}
	\caption{(Color online) The degree of boson excitation $\langle \hat{S}_z \rangle$  in the  S/AS basis (or, equivalently, the degree of coherence between the L and R wells) in the ground state and plotted as a function of $W$. Note the change in behaviour at the critical point $W=W_c= 0.5J$. The solid black curve shows the $N \rightarrow \infty$ case
         and the red dashed curves represent different values of $N$: 2,
         4, 6, 8, 10, where the arrow indicates increasing $N$. The values of the other parameters are $\Delta
         \epsilon = \Delta \epsilon^a = 0$, $J^a = J$. According to Eq.~(\ref{eq:Zposition}) $\langle \hat{S}_z \rangle \rightarrow 0$ as $W \rightarrow \infty$.}
	\label{fig:SZ}
\end{figure}

\begin{figure}[t]
\begin{center}
\includegraphics[width=1.0\columnwidth]{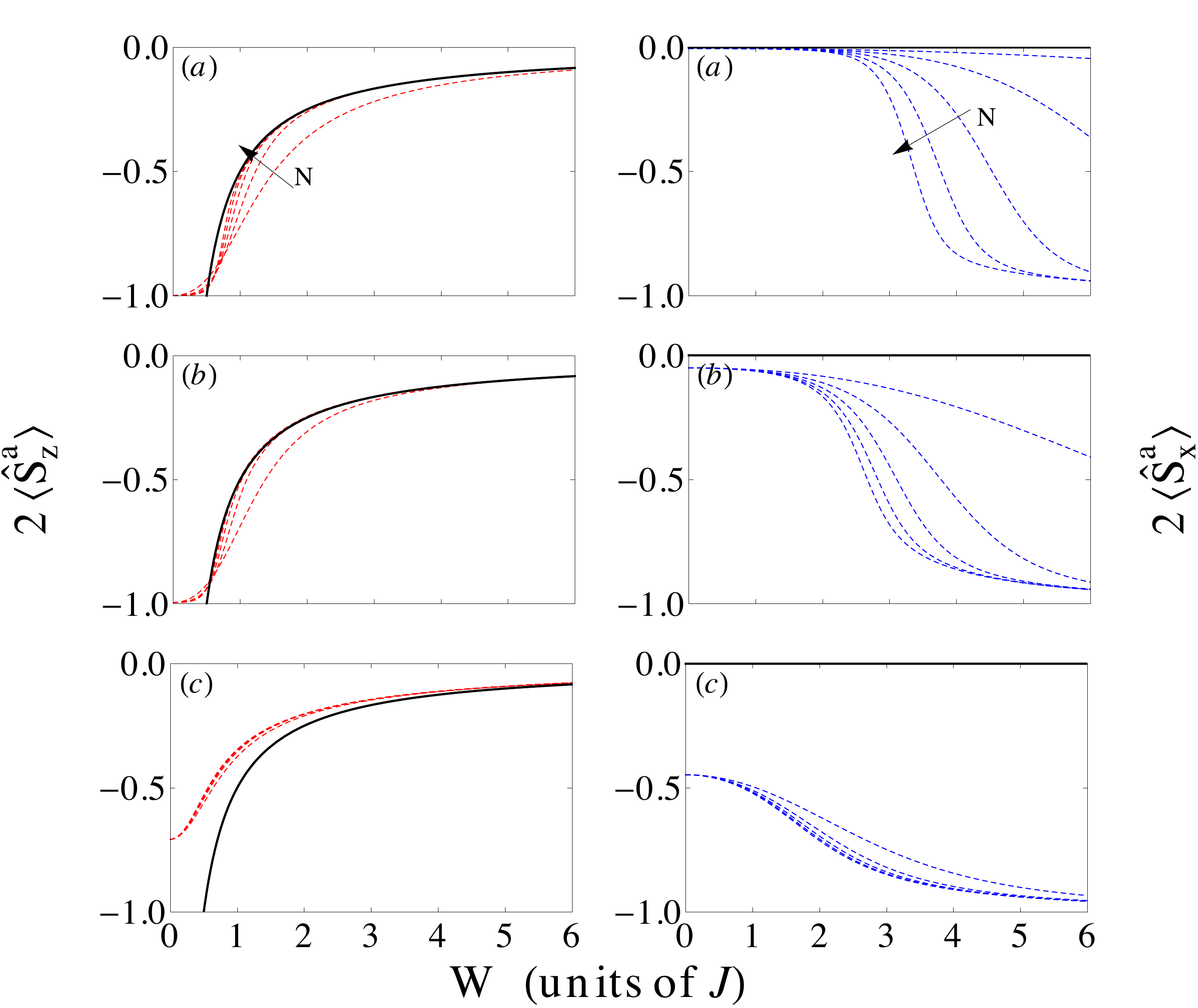}
\end{center}
	\caption{(Color online) The effect of a finite boson tilt upon the impurity.
	The left hand column shows  $\langle \hat{S}^a_z \rangle$ as a function of $W$ for the ground state which measures the degree of  excitation from the S mode into the AS mode (or, equivalently, the coherence between the L and R modes). The right hand column shows  $\langle \hat{S}^a_{x} \rangle$ which is the conjugate quantity to $\langle \hat{S}^a_z \rangle$. Each row is
          for a different value of the boson tilt: (a) $\Delta \epsilon = 0.01J$,
          (b) $\Delta \epsilon = 0.1J$, and (c) $\Delta \epsilon =J$.  The dotted lines are each for a different $N$: 2, 4, 6, 8,
          10, where the arrow indicates the direction of increasing $N$. The solid black curve plots the $N \rightarrow \infty$ limit
          for \emph{zero tilt}, i.e.\ $\Delta \epsilon = 0$. The values of the other parameters are $J^a = J$, $W_c
       = 0.5J$, $\Delta \epsilon^{a}=0$.}
	\label{fig:Sztilt}
\end{figure}

When $W>W_c$ the limiting case is   
\begin{equation}
\lim_{N \to +\infty} \frac{2 \langle \hat{S}_z \rangle}{N} = - \frac{J}{2 W}
\sqrt{\frac{J^{a^2}+4W^2}{J^2+4W^2}} \, .
\label{eq:Zposition} 
\end{equation}
Figure \ref{fig:SZ} displays $\langle \hat{S}_z \rangle$ as function of
$W$ for different values of $N$.  The solid black curve gives the thermodynamic limit and was calculated using the mean-field theory, and the dashed red curves are each for a different value of $N$ and were calculated using the full quantum theory. It is perhaps surprising how small
a number of bosons is needed in order to converge to the $N \rightarrow \infty$ limiting case, which agrees with the fact that the model becomes critical in the thermodynamic limit.  From Fig.\ \ref{fig:SZ} we see that when $W<W_c$ the bulk of the
bosons remain in the symmetric mode, and for $W>W_c$ there is a
macroscopic excitation into the antisymmetric mode tending to an equal population as $W \rightarrow \infty$, much like the infinite temperature limit for spins in a magnetic field where the populations of spin up and down become equalized.  The  Dicke model shows very similar qualitative dependence on $g$~\cite{emary03}, the difference being the sensitivity: the Dicke model has more excitations for equal values of $g$ and $W$. This is because the two-state nature of the impurity means that it can saturate, unlike a harmonic oscillator. Thus, when $W \rightarrow \infty$ the excitation of the impurity asymptotes to $\langle \hat{S}_{z}^{a} \rangle =0$  (i.e.\ $Y=\pm 1/2$), whilst the number of photons in the Dicke model
increases quadratically for large values of $g$
\begin{equation}
\lim_{N \to +\infty} \frac{2 \langle \hat{c}^{\dagger} \hat{c}
  \rangle}{N} = \frac{2 g^2}{\omega_A^2} \left ( 1 - \frac{\omega_A^2
    \omega_B^2}{16 g^4}  \right )\, .
\end{equation}    
The degree of excitation of the impurity and the number of photons will only be of the same order near the critical value of $g$.

\begin{figure*}[t]
\begin{center}
\begin{tabular}{c}
\includegraphics[width=2.0\columnwidth]{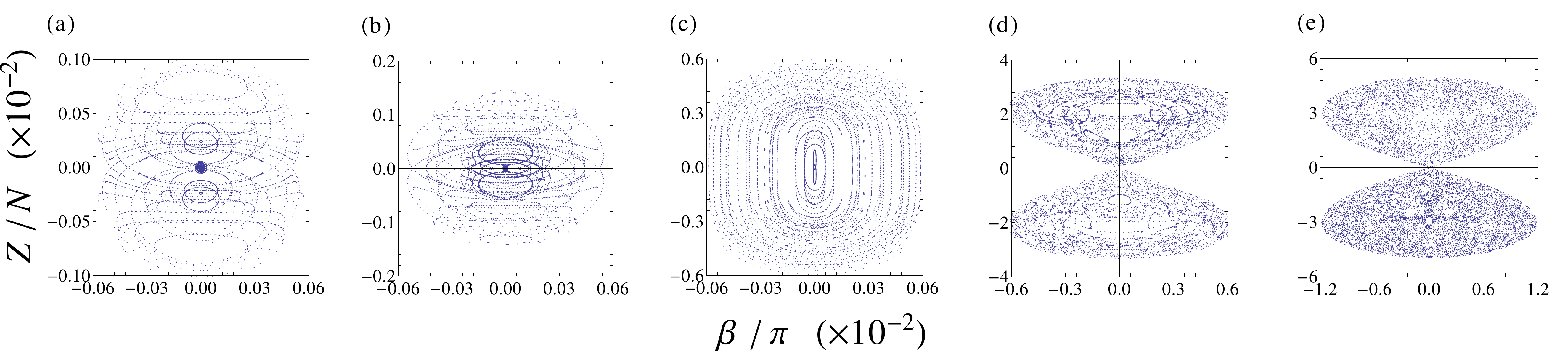} \\
\includegraphics[width=2.0\columnwidth]{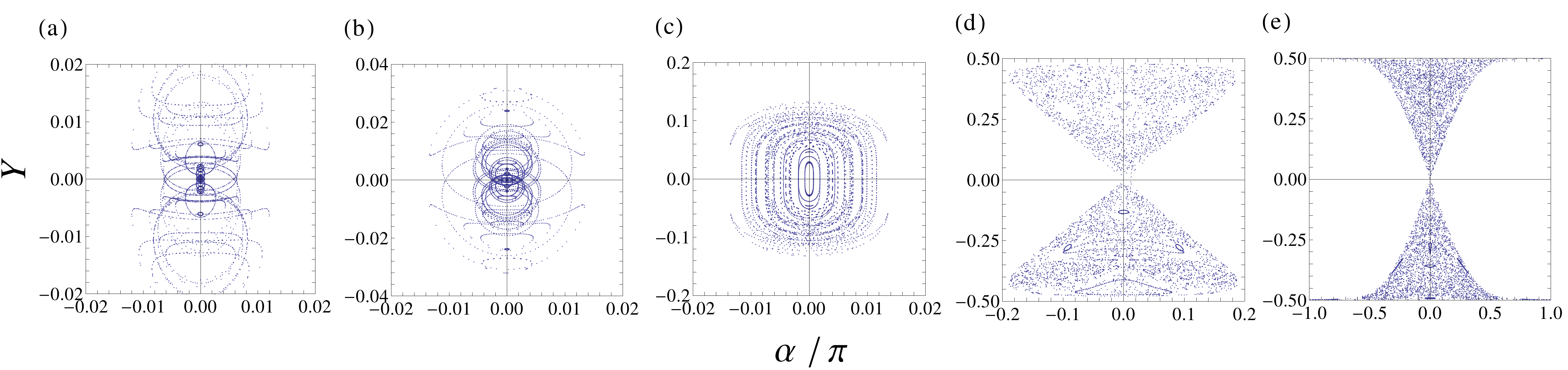}
\end{tabular}
\end{center}
	\caption{(Color online) Poincar\'{e} sections showing the emergence of chaos as $W$ increases through $W_c$.  For each plot,
          thirty random on-shell initial conditions are used and are
          evolved over a period $\tau=150$ ($\tau=Jt/\hbar$).  Each
          row shows intersections in different 2D planes: the top row is
          the $\beta Z$ plane with each point corresponding to $\alpha
          = 0$ and the bottom row is the $\alpha Y$ plane with each
          point corresponding to $\beta = 0$.  From left to right the
          plots increase in $W$: (a) $W = 0.50W_{c}$, (b) $W = 0.75W_{c}$,
          (c) $W = W_{c}$, (d) $W = 1.25W_{c}$, and (e) $W = 1.50W_{c}$.  The values of the other
          parameters are $N=500$, $\Delta \epsilon=\Delta
          \epsilon^{a}=0$, $J^a = J$ and $E_{\mathrm{shell}} =
          -501J$. Please note the different ranges for each panel.}
	\label{fig:PPanel}
\end{figure*}

So far we have dealt with the idealized case of no asymmetries (tilt) in the
double-well. A non-zero tilt breaks the $Z_2$ parity symmetry and this prevents the system from being critical in the thermodynamic limit. The corresponding effect in the Dicke model is obtained by driving the boson mode and/or the spins~\cite{jonas3}. In Fig.\ \ref{fig:Sztilt} we show the effect of different boson tilt values on the impurity by plotting $\langle \hat{S}_z^a \rangle$ and $\langle \hat{S}_x^a \rangle$ as functions of $W$. Analogously to the equivalent quantities for the bosons, the interpretation of $\langle \hat{S}_z^a \rangle$ is that it gives the degree of excitation (number difference) of the impurity from the S mode into the AS mode, or, equivalently, the degree of coherence of the impurity between the L and R modes which vanishes when the impurity settles into just one well (as determined by the applied boson tilt).  $\langle \hat{S}_x^a \rangle$ is the conjugate quantity and gives the coherence between the S and AS modes, or, equivalently,  the number difference between the L and R modes.
We see that the tilt does not have a great effect in the vicinity
of $W_c$ until panel (c), where the tilt has the same
magnitude as the tunnelling energy and the system ``realizes'' it is tilted. To understand this better we note that in the non-interacting limit, $W=0$, the hamiltonian in the S/AS basis for the impurity is simply $\hat{H}_\mathrm{I}=2J^a\hat{S}_z^a-\Delta\epsilon^a\hat{S}_x^a$. Hence,  for $W=0$   we have $\langle\hat{S}_z^a\rangle=-J^a/\sqrt{\left(4J^a\right)^2+\left(\Delta\epsilon^a\right)^2}$. This finite degree of S/AS excitation, or equivalently, reduction in L/R coherence even in the non-interacting regime is yet another sign that for a non-vanishing tilt of the double-well the criticality appearing in the thermodynamic limit is lost. Indeed, the finite tilt restores the correspondence between the mean-field and quantum results: without a tilt the quantum system does not choose a particular well and enters a Schr\"{o}dinger cat state which has no classical correspondence. Only when fluctuations due to an external environment are included does the cat state collapse randomly to one or other of the two wells. On the other hand, in the presence of tilt the cat state never really forms.

\section{Mean-field Dynamics}
\label{sec:mfc}
The main feature of classical dynamics in the Dicke model is global
chaos when $g > g_c$ \cite{altland12}.  Here we use Poincar\'{e} sections through phase space as a
tool to investigate the emergence of chaos in the
boson/impurity system.  Fig.\ \ref{fig:PPanel} shows Poincar\'{e} sections for the
bosons (top row) and the impurity (bottom row) as $W$ is increased.
The dynamics take place on an energy shell with energy equal to that of the
unstable point located at $(0,0,0,0)$, $E_{\mathrm{shell}} = -NJ - J^a$.  This corresponds to the center of the lowest band, and when the loop appears it becomes the upper branch. We see that as $W$ increases chaos emerges and for $W>W_c$ chaos is
dominant and ergodicity is observed. In Appendix \ref{sec:app} we
show Poincar\'{e} sections for a fixed value of $W$ above $W_c$ on
different energy shells in order see how chaos depends on our position in the spectrum.

\begin{figure}[t]
\begin{center}
\includegraphics[width=0.8\columnwidth]{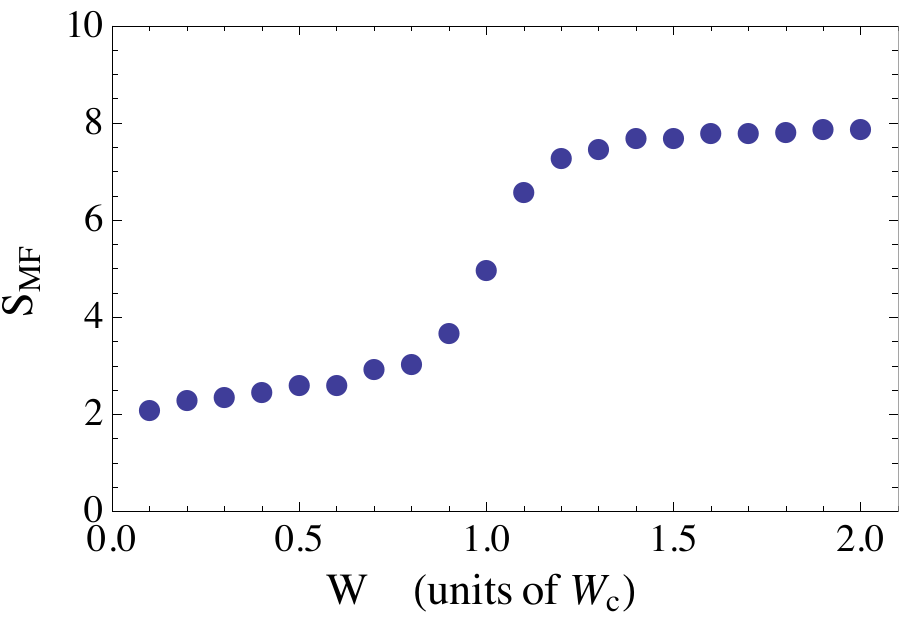}
\end{center}
	\caption{(Color online) Mean-field entropy as a function of $W$ showing a jump at $W_c$.  Each point
          is calculated using Eq.\ (\ref{eq:entropy}) after dividing 
          Poincar\'{e} sections for the impurity into subintervals
          $10^{-2}$ times the size of the $\alpha Y$ plane.  The other parameters of the plot are
$J^a = J$, $N= 500$, $\Delta \epsilon=\Delta
          \epsilon^{a}=0$, and $E_{\text{shell}} = -501J$.   For
each calculation forty random on-shell initial conditions were
used and plotted until 3,000 intersections with the Poincar\'{e} plane were produced.}  
	\label{fig:MFE}

\end{figure}

In order to locate the precise value of $W$ at which chaos emerges we divide the impurity phase
space into subintervals defining a probability as $p_i = m_i/ \textrm{M}$ where
$\textrm{M}$ is the number of subdivisions and $m_i$ is the number of points in
the $i^{\text{th}}$ subinterval.  With this probability we can define
an entropy in the usual way 
\begin{equation}
S_{\mathrm{MF}} = - \sum_{i} p_i \,  \textrm{ln} \, p_i \, .
\label{eq:entropy}
\end{equation}
Equation (\ref{eq:entropy}) can be thought of as a way to quantify the
area of the phase space the impurity can explore. Even if the value of $S_\mathrm{MF}$ depends quantitatively on the partitioning of the phase space, we expect it to show some generic qualitative features. For example looking at both
extremes, if the points can be found entirely in one subinterval, then
$S_{\mathrm{MF}} = 0$ and if the points are maximally spread, then
$p_i = 1/ \textrm{M}$,
for all $i$, and $S_{\mathrm{MF}} = \textrm{ln} \, \textrm{M}$.  Since a system becomes
ergodic when chaos is dominant we expect higher values of $S_{\mathrm{MF}}$ as
$W$ increases.  Looking at Fig.\ \ref{fig:MFE} this is exactly what we
find.  At $W = W_c$ there is a jump in $S_{\mathrm{MF}}$ signalling the onset of
ergodicity.

\begin{figure}[t]
\begin{center}
\includegraphics[width=1.0\columnwidth]{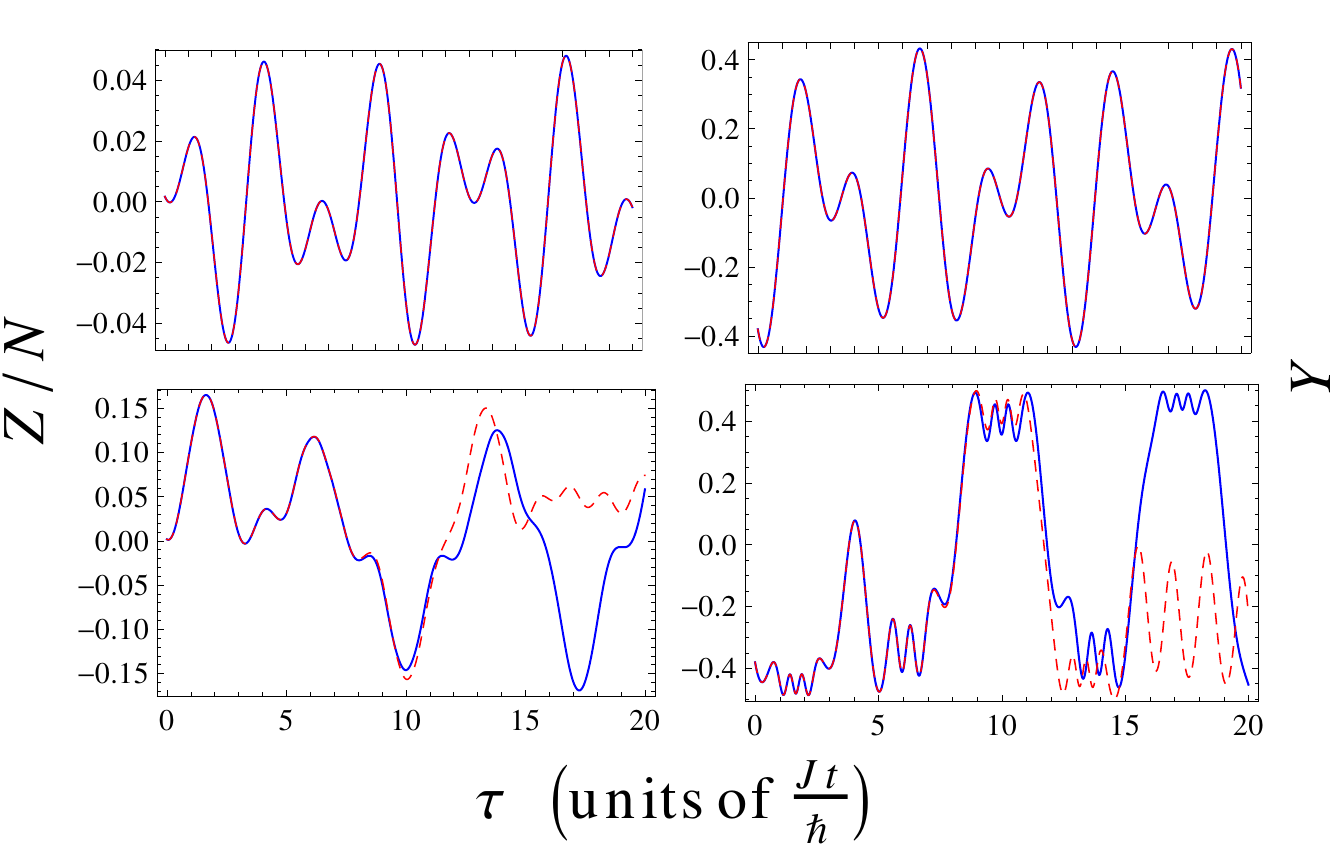}
\end{center}
	\caption{(Color online) The atom
          number difference between the right and left wells as a
          function of the dimensionless time parameter $\tau$ ($\tau =
          Jt/\hbar)$.  Each row corresponds to a different value of
          $W$: $W = 0.5W_{c}$ (top row) and $W = 2.0 \,
          W_{c}$ (bottom row).  Each plot is generated using the same initial
          conditions for the number differences for the bosons and
          the impurity on the energy shell $E_{\mathrm{shell}} = 100.5J$.  The dashed red curves and solid blue curves differ
          by $\Delta Z/Z = \Delta Y/Y = 10^{-4}$.  The other
          parameter values are $N = 100$, $J^a = J$, and $\Delta \epsilon=\Delta
          \epsilon^{a}=0$.}
	\label{fig:trajcomb}
\end{figure}

Next, we look at the most common aspect of classical chaos---sensitivity to initial conditions.  Figure \ref{fig:trajcomb} shows the
time dependence of $Y$ and $Z$ for $W<W_c$ (top row) and $W>W_c$
(bottom row).  All dynamics take place on $E_{\mathrm{shell}} = -NJ-J^a$  with
two trajectories initially separated by $\Delta Z/Z = \Delta Y/Y = 10^{-4}$.  For
$W<W_c$ we see that both trajectories remain close.  However, for $W>W_c$
we see the trajectories begin to diverge at $\tau \approx 10$
signalling a loss of information about the initial state of the
system.

\section{Impurity induced self-trapping}\label{sec:selft}
Generally speaking, classical trajectories that set off in the vicinity of a stable fixed point remain close to the fixed point. Thus, the pitchfork structure of the fixed points, as demonstrated in Fig.~\ref{fig:pitchforkz}, implies that the classical system can become locked with a large population imbalance of the bosons---a large fraction of the bosons remains in one well and does not tunnel to the other well. This is the phenomenon of self-trapping~\cite{milburn97,smerzi97}. Self-trapping in bosonic Josephson junctions derives from the self-interaction between the atoms and can maintain large differences in the populations of the two wells. Roughly speaking, the interaction effectively shifts the onsite energies in the two wells, and whenever there is a large population imbalance and strong shifts the coherent tunneling becomes heavily detuned which therefore hinders the oscillations. While this effect is rather general, for atomic condensates it was first demonstrated in a BEC double-well system~\cite{albiez05}. 

\begin{figure}
\begin{center}
\includegraphics[width=1.0\columnwidth]{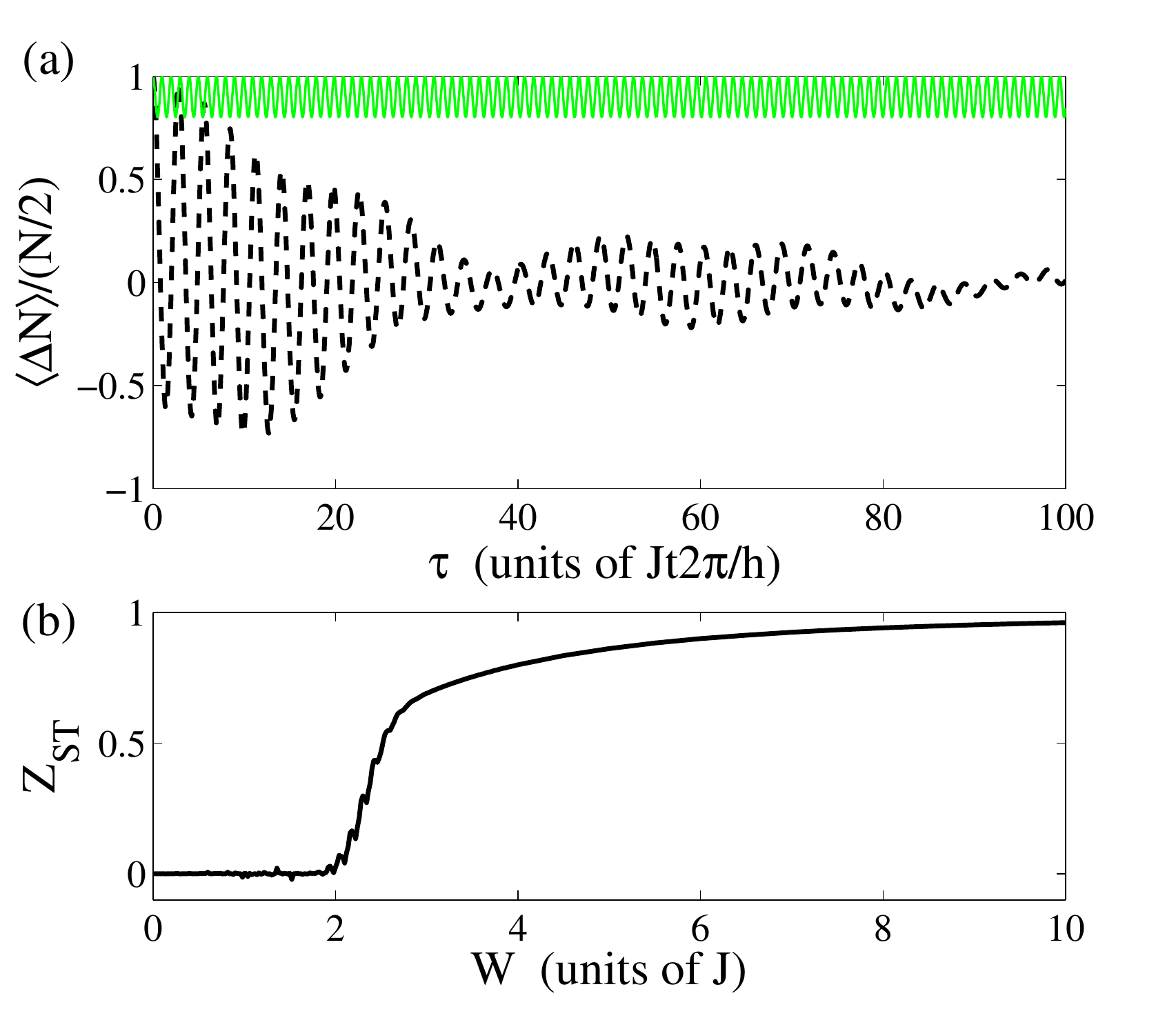}
\end{center}
	\caption{(Color online) The evolution of the scaled boson
          population imbalance (a), and the long-time time average of the imbalance (b). In the upper plot, the dashed black curve displays the time evolution for an interaction strength $W=1$. For this interaction strength, no self-trapping occurs and the collapse of oscillations are due to the build-up of impurity-boson correlations. For $W=6$ (green solid curve) self-trapping is clearly apparent. Plot (b) demonstrates how the self-trapping sets in at around $W\approx2$ for the current parameters. The initial state has all the bosons in the right well and the impurity in the left well, and the rest of the parameters are $J^a=J$, $\Delta\epsilon=\Delta\epsilon^a=0$, and $N=100$.}
	\label{fig:selft}

\end{figure}

The situation is different in the present set-up where the bosons are non-interacting and so self-trapping can only stem from the boson-impurity interaction. A basic understanding of this case can be gained by fixing a value of $\Delta M \neq0$ and setting $\Delta\epsilon=\Delta\epsilon^a=0$. Referring to the hamiltonian (\ref{eq:mbham}) expressed in the L/R basis we see that as far as the bosons are concerned the impurity acts as an effective tilt which is the origin of the self-trapping. In this `adiabatic' picture the motion of the bosons is free and can solved exactly. In a complete description the state of the impurity atom is itself also evolving and the coupled boson-impurity dynamics must be taken into account.

We demonstrate self-trapping by integrating the full quantum model of Eq.\ (\ref{eq:mbham}) for an initial state of $N$ bosons in the right well and the impurity atom in the left well~\cite{com}. For small interactions $W$, both the impurity and the bosons display coherent oscillations between the two wells, as shown by the dashed black curve of Fig.~\ref{fig:selft} (a). The mixing of time-scales in this regime leads to a relaxation of the oscillations. During the decay period a large entanglement is shared between the impurity and the bosons. Increasing $W$ now leads to a rapid decrease of the amplitude of the Josephson oscillations in agreement with the expected trapping effect (green solid line). An estimate of the degree of the self-trapping can be obtained by calculating the long-time time average
\begin{equation}
Z_\mathrm{ST}=\frac{1}{T_2-T_1}\int_{T_1}^{T_2}dt\,\frac{\langle\Delta\hat{M}\rangle}{N/2},
\end{equation}
where $T_2\gg T_1\gg0$ are two long times (as discussed below, there is another time-scale for which the self-trapping is lost, and $T_1$ and $T_2$ should be long compared to the Josephson oscillation period but short compared to the decay of the trapping effect). Figure \ref{fig:selft} (b) shows the $W$-dependence of $Z_\mathrm{ST}$. There is a sudden onset of self-trapping at around $W\approx2$ for which the population imbalance increases rapidly and tends asymptotically to 1. We have numerically determined that the critical interaction $W_\mathrm{st}$ for which self-trapping starts is only weakly dependent of atom number $N$ and $J^a$, while it scales linearly with $J$, more precisely  $W_\mathrm{st}\approx2J$. Thus, the onset of self-trapping does not appear exactly at the critical coupling $W_c$ for the bifurcation.

\begin{figure}
\begin{center}
\includegraphics[width=1.0\columnwidth]{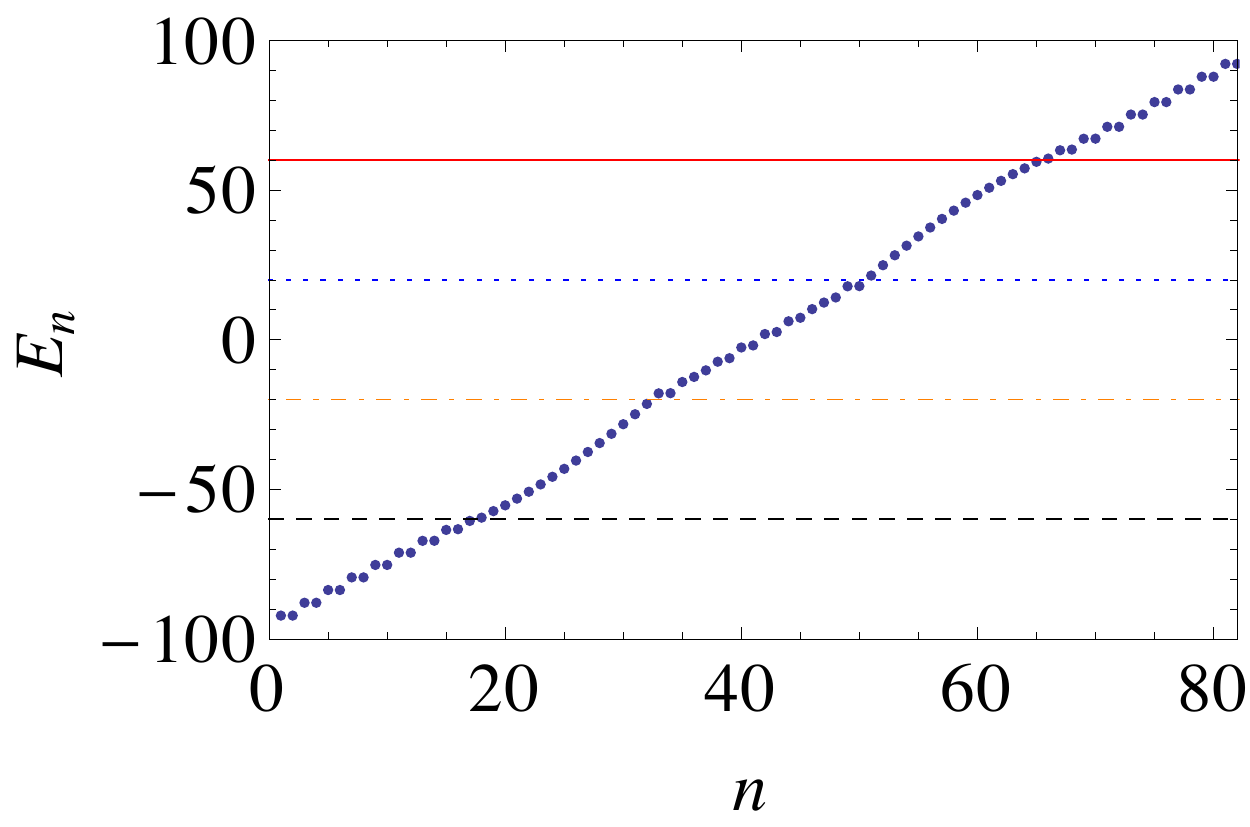}
\end{center}
	\caption{The quantum energy levels (blue dots) for $N = 40$,  $J^{a} = 20 J$, $W = 4 J$, $W_{c} = \sqrt{2} J$ and $\Delta \epsilon =\Delta \epsilon^{a}=0$. We have reduced $N$ in this figure so that the finer details of the spectrum are visible. The four horizontal lines give the positions of the mean-field stationary solutions as shown in Fig.\ \ref{fig:EvTilt} and we have maintained the same color scheme as there, namely, in ascending order: $E=-NJ-J^{a}$ ($\alpha=\beta=0$, dashed black), $E=-NJ+J^{a}$ ($\alpha=\pi,\beta=0$, dot-dashed orange), $E=NJ-J^{a}$ ($\alpha=0,\beta=\pi$, dotted blue), $E=NJ+J^{a}$ ($\alpha=\beta=\pi$, solid red).  Note that because $W>W_{c}$ the lowest and highest energy classical solutions have loops and the positions given here correspond to the unstable branch, that is the highest and lowest branch, respectively.}
	\label{fig:specSTsection}
\end{figure}

As already mentioned, the Hamiltonian (\ref{eq:mbham}) supports a $Z_2$ parity symmetry. Each parity sector constitutes a separate spectrum and for non-zero $W$ and in the large $N$ limit the two spectra become identical. The energy gap $\delta$ between corresponding even and odd parity eigenstates is found to close exponentially fast with $N$, i.e. $\delta\sim\exp(-aN)$ for some $N$-independent constant $N$. In the self-trapping regime the gap also closes exponentially with the interaction strength $W$, an effect that is associated with below-barrier tunnelling in the double well potentials that form in Fock space when $W>W_{c}$ and whose stationary points give the pitchfork bifurcation. To illustrate this the full energy spectrum is plotted in Fig.\ \ref{fig:specSTsection} for parameters within the self-trapping regime, i.e.\ for $W>2W_{c}$.  The most striking feature of 
Fig.\ \ref{fig:specSTsection} is that inside the lower and upper loops \emph{all} the energy levels are paired up in quasi-degenerate pairs.

To see how self-trapping works schematically consider a state initially localized in the L well. This state can be made by the superposition of an even parity eigenstate and an odd parity eigenstate $\vert \mathrm{L} \rangle= (\vert \mathrm{E} \rangle \exp[-i \omega_{\mathrm{E}}t]+ \vert \mathrm{O} \rangle \exp[-i \omega_{\mathrm{O}}t])/\sqrt{2}$. If these states make up one of the quasi-degenerate pairs  then the difference in the two energies is exponentially small $\delta=\hbar \omega_{\mathrm{O}}-\hbar \omega_{\mathrm{E}}$ and the time evolution of the superposition into the $\vert \mathrm{R} \rangle$ state is very slow. We have verified this numerically and also that the characteristic time for this collapse scales as $\tau_\mathrm{col}\sim K^N$ for some constant $K$. Naturally, this exponential growth of the collapse time means that the collapse will most likely be far beyond any realistic experimental observation.  A general self-trapped state will be a projection over many eigenstates but if the wave packet lies entirely within one of the loops it will be self-trapped because it will be solely made up of quasi-degenerate even and odd pairs. However, the energy separation between different pairs introduces a different and larger energy scale than the tunnel splitting effect and hence a faster oscillation about the mean value of $\Delta N$ as can be seen in the solid green curve in Fig.~\ref{fig:selft} (a).

\begin{figure}
\begin{center}
\includegraphics[width=1.0\columnwidth]{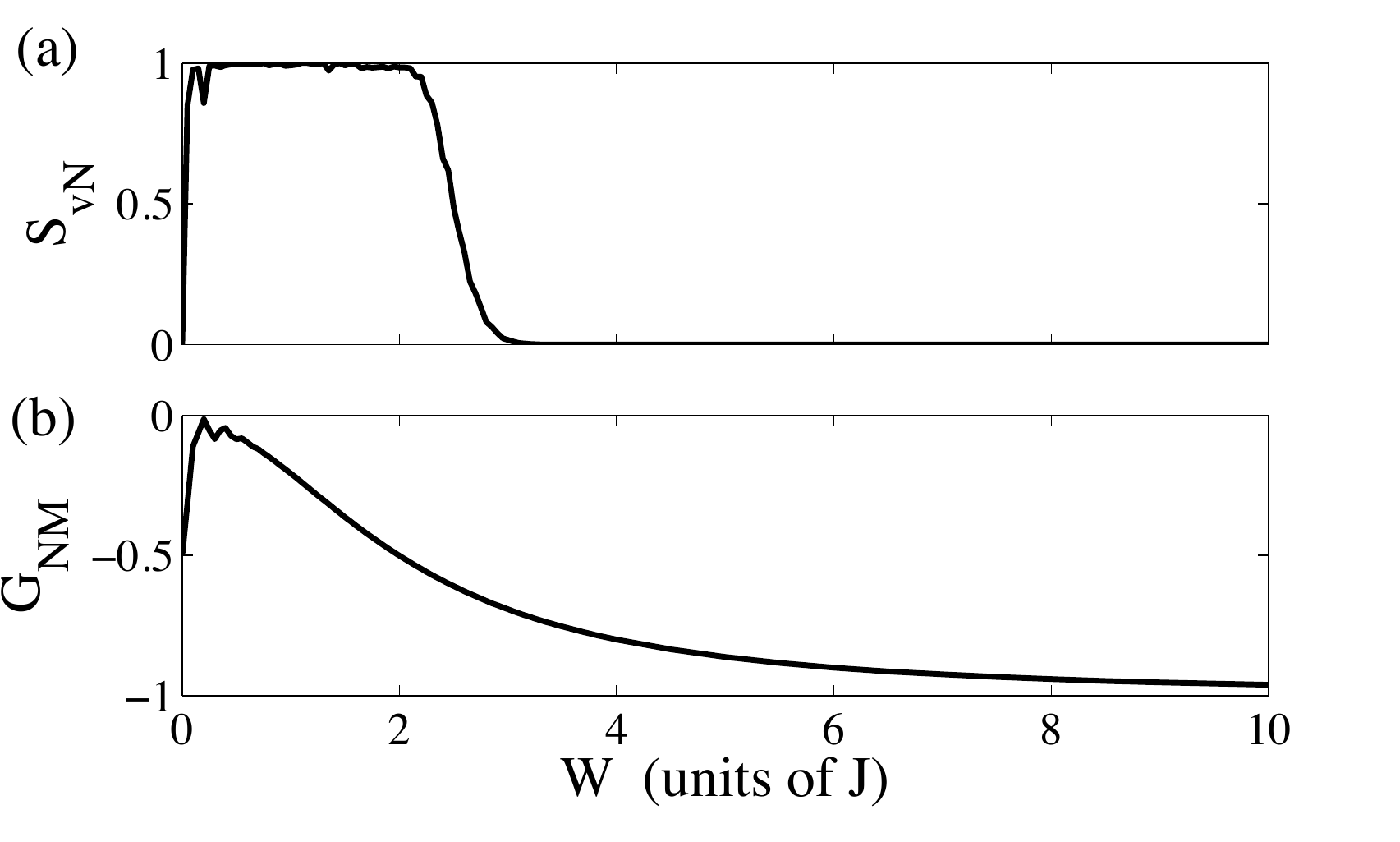}
\end{center}
	\caption{The time-averaged von Neumann entropy $S_\mathrm{vN}$ (a) and scaled correlator $\langle\Delta\hat{N}\Delta\hat{M}\rangle/(N/2)$ (b). The two plots demonstrate the absence of entanglement and presence of classical anti-correlations deep in the self-trapping regime. The initial state and the parameters are the same as in Fig.~\ref{fig:selft}.}
	\label{fig:ent}

\end{figure}

It has been argued that the mechanism behind self-trapping in a bosonic Josephson junction can be viewed as a quantum Zeno effect; the atom-atom interaction acts as an effective measurement in which the state of single atoms is measured by the remaining ones~\cite{zeno1} (external measurement induced Zeno effects on self-trapping have also been discussed~\cite{zeno2}). The question then arises as to whether in the present self-trapping set-up the bosons perform an effective measurement on the impurity  (or vice versa)? During a standard quantum measurement the meter (e.g.\ the bosons) becomes entangled with the system (e.g.\ the impurity) and in order for the measurement to distinguish between the two possible states they should be macroscopically distinguishable, i.e.\  a Schr\"{o}dinger cat state should form. Finally, the cat state is collapsed by an environment and the state of the system can be read off from the state of the meter with which it is now perfectly classically correlated. The case of self-trapping is different because deep in the self-trapping regime the bi-partite state factorizes as the bosons and the impurity occupy \emph{definite} positions due to our initial preparation of the system and are not in superpositions of both wells. In other words they are classically anti-correlated but there is approximately no quantum entanglement shared between the two parties, i.e.\ the von Neumann entropy $S_\mathrm{vN}=-\mathrm{Tr}_\mathrm{im}\left[\hat{\rho}_\mathrm{im}\log(\hat{\rho}_\mathrm{im})\right]$ (where $\hat{\rho}_\mathrm{im}$ is the reduced density operator for the impurity atom, the trace is over the bosons, and the logarithm is to base two) approaches zero in the self-trapping regime, while the correlator $G_\mathrm{NM}=\langle\Delta\hat{N}\Delta\hat{M}\rangle/(N/2)$ goes towards -1. The above arguments are demonstrated in Fig.~\ref{fig:ent}; (a) shows the time-averaged entropy and (b) the time-averaged correlator. It is noteworthy that the convergence of the correlator is slower than that of the entropy. We have not ruled out the quantum Zeno effect being at work here because that too begins with evolution from a known initial state. More discussion of the von Neumann entropy for this system and particularly the effect of tilt can be found in Section VIII of \cite{mulansky11}.
 As a final remark we note a related phenomenon in quantum optics, namely {\it population trapping}~\cite{pt}. This effect occurs as frozen population transfer between internal atomic states even in the presence of a drive.

\section{Level-spacing distribution}
\label{sec:ls}
It is a remarkable fact that the spectra of quantum systems whose classical limit is chaotic are statistically distinct from those whose classical limit is regular~\cite{haake00}.  Somewhat counter-intuitively, the energy levels of classically regular systems are distributed randomly (providing symmetries that cause degeneracies are removed) so that the probability distribution for the spacings $S$ between neighbouring energy levels is Poissonian
\begin{equation}
P_{\mathrm{P}}(S) = e^{-S} \, .
\end{equation}
Conversely, the energy levels of classically chaotic systems are correlated with each other and display level repulsion  \cite{gubin12,graham87,brown08,stone10}.  The probability distribution in the chaotic case can be inferred through the study of
random matrices \cite{haake00,wigner51,dyson70}. The boson-impurity system is expected to display a level
spacing distribution closest to the Gaussian orthogonal ensemble
(GOE) since this ensemble represents real
symmetric matrices.  The GOE obeys the  Wigner-Dyson distribution 
\begin{equation}
P_{\mathrm{WD}}(S) \approx \frac{\pi S}{2} \, e^{- \frac{\pi S^{2}}{4}} \, .
\label{eq:WD}
\end{equation}
However, the elements of a GOE matrix are chosen
randomly from a Gaussian distribution, whereas most of the elements 
in our Hamiltonian are zero (in the Fock basis it is tridiagonal).  Physically this means that the GOE describes
systems with infinite range interactions \cite{santos12} whereas the boson-impurity system
has short range ones.  Therefore, we should not expect our
system to display all the properties of a GOE system.


The first step in obtaining the level spacing distribution is to
separate the eigenvalues based on their symmetries.  The reason for
this is because symmetries cause non-generic features such as the degeneracies we find in our Hamiltonian due to the $Z_{2}$ parity symmetry.
As already pointed out, $H$ conserves the number of particles.  This symmetry breaks up the hamiltonian into independent blocks, each of dimension $2(L - 1 +N)!/[(L-1)!N!]$ \cite{gubin12} where $N$ is the total number of particles, $L$ is
the number of wells and the factor of two comes from the impurity.
Since we have a double well, the size of each block is $2N+2$ which is
the whole hamiltonian.  Therefore, we can conclude that particle
number conservation does not affect the energy level spacings.  The parity of states is also conserved which causes the
hamiltonian to be broken up into two (even/odd parity) blocks of dimension $N+1$.
We perform statistics on the two parity blocks separately and add the results together at the end.  

\begin{figure}
\begin{center}
\includegraphics[width=0.8\columnwidth]{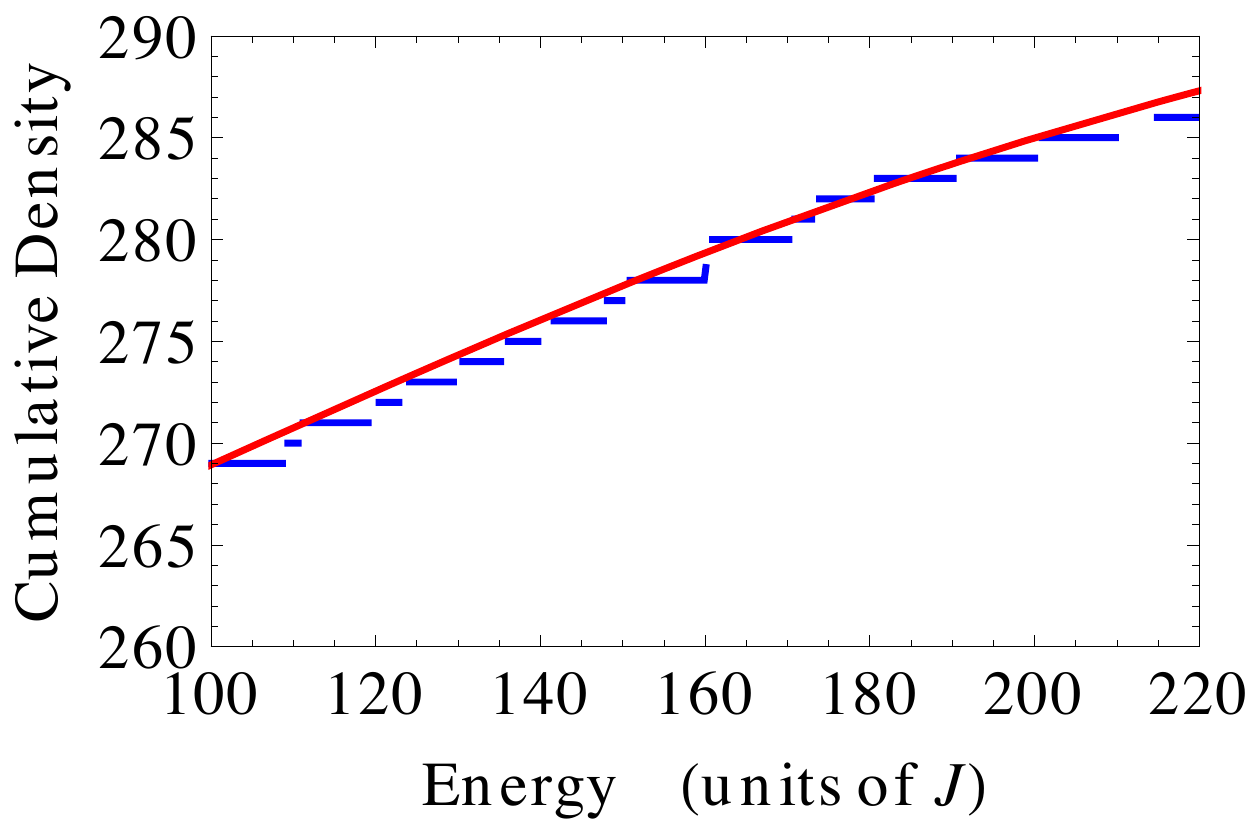}
\end{center}
	\caption{(Color online) A sample of $N(E)$, blue steps, and
          $\bar{N}(E)$, red line.  Parameter values are $N=500$, $J
          =1.282756$, $J^a = 466.4946$, $W \approx 2.29W_c$, and $\Delta \epsilon=\Delta \epsilon^{a}=0$.  }  
	\label{fig:cumuden}

\end{figure}

The second step in obtaining the level spacing distribution to \textit{unfold} the spectrum of each block.  This process
rescales the local mean level spacing so that it is equal to unity, allowing spectra from
different systems or also different regions of the same spectrum to
be compared.  To visualize the regularity of the level spacing and to better
understand the unfolding process we examine the cumulative density
\begin{equation}
N(E) \equiv \sum_{i} \theta (E - E_i)
\end{equation}
where $\theta$ is the Heaviside step function. $N(E)$ is the number of energy levels with energy less than $E$ and can be thought of as having
a smooth part $\bar{N}(E)$, and a fluctuating part 
$N_{\mathrm{fl}}(E)$, 
\begin{equation}
N(E) = \bar{N}(E) + N_{\mathrm{fl}}(E)
\end{equation}
 The form of $N(E)$ and
$\bar{N}(E)$ can be seen in Fig.\ \ref{fig:cumuden} where $N(E)$ is the staircase
function (blue curve) and $\bar{N}(E)$ is the smooth function (red curve). The process of
unfolding amounts to subtracting the smoothened part of $N(E)$ and keeping
the fluctuations.  To do this we use the map $E \rightarrow x$ \cite{haake00}, so
that
\begin{equation}
x_i = \bar{N}(E_i); \hspace{5mm} i=1,2,...,N \, ,
\end{equation}
where $\bar{N}(E)$ is calculated by fitting the chaotic region of the spectrum to a
seventh degree polynomial \cite{stone10}.  The level-spacings, $S$, can now
be calculated as
\begin{equation}
S_i = x_{i+1} - x_i; \hspace{5mm} i=1,2,...,N-1 \, .
\end{equation}

\begin{figure}[h]
\begin{center}
\includegraphics[width=0.9\columnwidth]{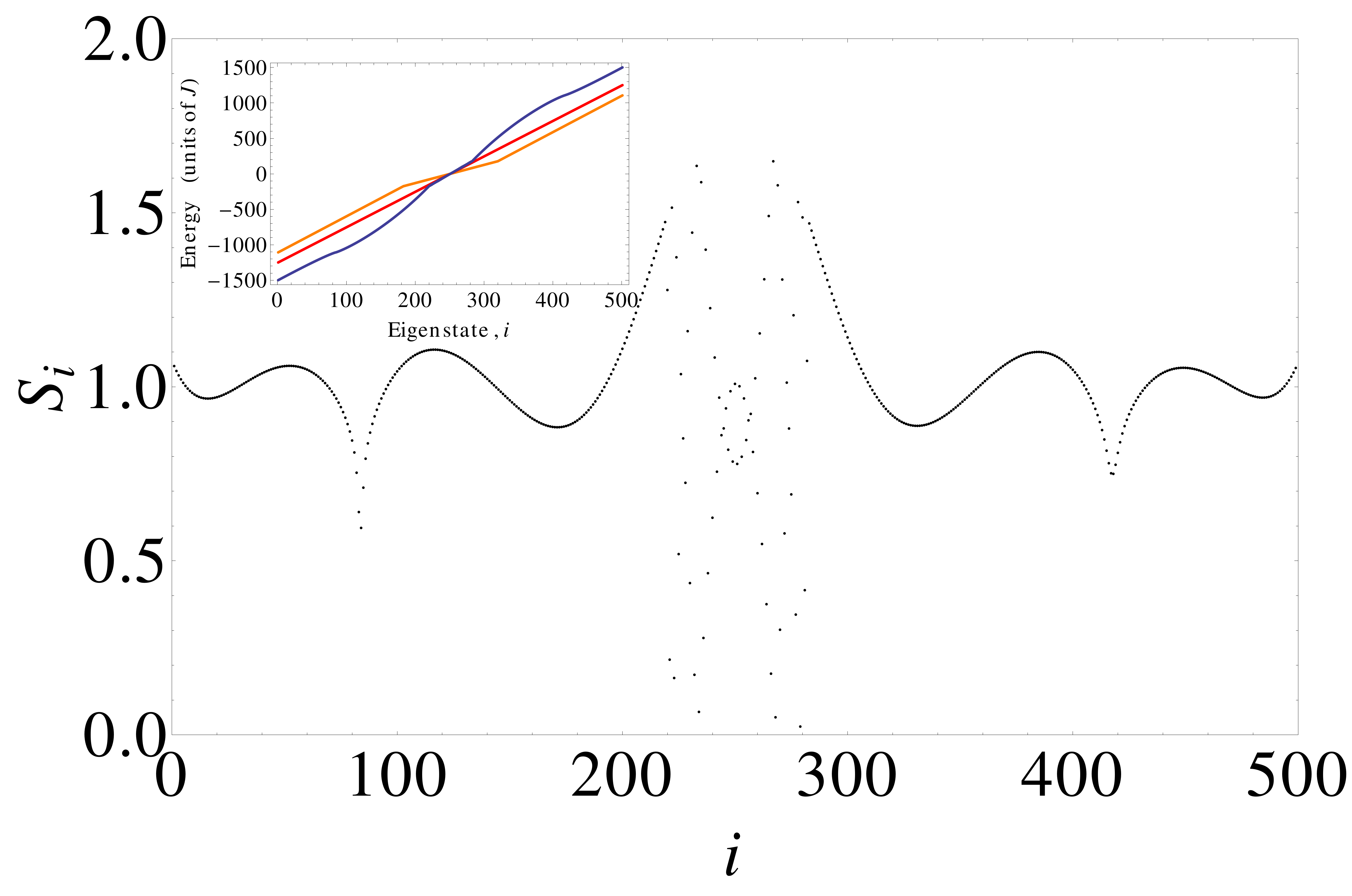}
\end{center}
	\caption{(Color online)  Level
          spacing as a function of the index for the blue spectrum in
          the inset.
          The inset shows energy spectra for different parameter values: $J =
          J^a = 0$ and $W = 5J$ (red), $J
          =1.282756$, $J^a = 466.4946$ and $W \approx 0.05W_c$
          (orange), $J=1.282756$, $J^a = 466.4946$ and $W
          \approx 2.29W_c$ (blue).
          The boson number is kept constant at $N=500$ and $\Delta \epsilon=\Delta \epsilon^{a}=0$.}
	\label{fig:ESpec}

\end{figure}

It is a rarity for systems to go from completely
regular to completely chaotic and instead are usually comprised of a
mixture of regular and chaotic regions as can be seen
in Fig.\ \ref{fig:PPanel}.  Indeed, the Kolmogorov-Arnold-Moser Theorem (KAM
Theorem) \cite{kolmogorov54,arnold61,moser62} properly explains why
regions of different stability gradually appear.  Our results in Appendix \ref{sec:app} show that different parts of the spectrum differ in their degree of irregularity and so we follow 
\cite{stone10} and only select the energy levels that overlap in the limits
where $W \ll J$ and $W \gg J$ because neither of the limits apply in
this region and these states are expected to be ``maximally
chaotic''.  We also choose $J$ and $J^a$ randomly such that the
maximum values of the three terms in the hamiltonian are $\mathcal{O}(N)$.  For
this section we use the values $J =1.282756$ and $J^a =
466.4946$.  In Fig.\ \ref{fig:ESpec},  we see some overlap in the
range $200 \leq n \leq 300$, and the inset shows $S_i$ as a function of
$i$.  Large fluctuations from the mean are clearly shown for spacings
of the states in the overlapping region.  The two dips in the inset are
from states with energies near $\pm(JN+J^a)$; these regions are
separatrix-like and have a divergence in the density of states causing
small energy spacings.

\begin{figure}[h]
\begin{center}
\includegraphics[width=1.0\columnwidth]{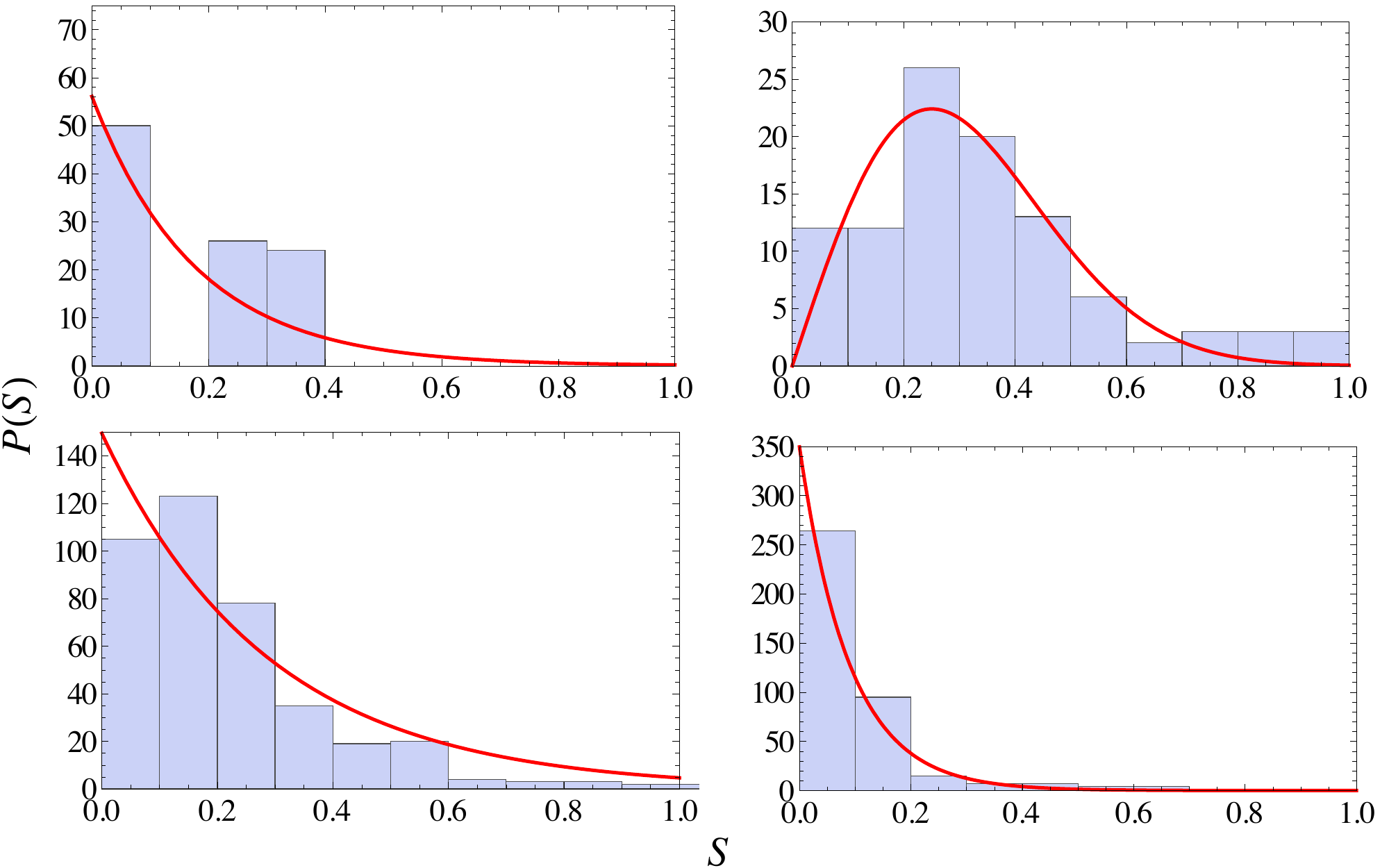}
\end{center}
	\caption{(Color online) Level-spacing distributions for $W \approx 0.05W_c$ (left
          column) and $W \approx 2.29W_c$ (right column).  The top
        row shows statistics for eigenstates $200 \leq n \leq 300$ and the bottom
       row for the remaining eigenstates. }
	\label{fig:lsgrid}

\end{figure}

Figure \ref{fig:lsgrid} shows the level-spacing distributions (after
the mean level-spacing is subtracted) for
different values of $W$.  Panels (a) and (b) show distributions for
the maximally chaotic states below and above $W_c$, respectively.
Panels (c) and (d) show the same for the remaining states.  We can see
from the lower row that the states we deemed least-chaotic resemble a Poissonian
distribution below and above $W_c$.  The reason panel (a) does not
resemble a Poissonian distribution is due to the fact that the
hamiltonian for $W \ll J$ essentially describes two decoupled 
oscillators with an approximate spectrum of
\begin{eqnarray}
E_{k,l} \approx J(2k-N)+J^a(2l-1); \hspace{2mm} && k=0,1,2,...,N
\nonumber \\
&& l=0,1 \, .
\end{eqnarray}
Groups of uncoupled harmonic oscillators usually have distributions of
staggered ``pillars'' due to their constant individual spacings \cite{emary03}.
Panel (b) resembles the Wigner-Dyson distribution and clearly shows an
increase in level repulsion as expected for a system which becomes
classically non-integrable and chaotic.  Of course, our system with a single impurity is only one step away from integrability (pure boson case) and so may not be irregular enough to see a clearer change from Poissonian to Wigner-Dyson distributions.
A more complex hamiltonian than Eq.\ (\ref{eq:mbham}) including, say, different intra-well interactions\cite{stone10}
or more impurities would most likely show a clear change.

\section{Conclusion and Discussion}
Adding a single impurity to a bosonic Josephson junction in a double well
can produce rich and interesting dynamics both at the mean-field level and in the many-body regime.
Using a stability analysis we studied swallowtail loops that emerge in the energy spectrum in the mean-field limit at a critical value $W_c$ of the interaction energy between the impurity and bosons. This critical value corresponds to a symmetry-breaking bifurcation in the ground state where there is a macroscopic re-organization of the bosons and it coincides with the critical coupling for a similar symmetry-breaking QPT in the Dicke model.

As $W$ is increased through $W_c$ we showed the emergence of classical chaos in
two ways: Poincar\'{e} plots showed fading of regular behaviour and an
increase in ergodicity; and trajectories with close initial conditions
remained close for $W < W_c$, but diverged for $W>W_c$. Complementary to the mean-field calculations, a statistical analysis of the quantum energy levels revealed level repulsion also sets in when $W>W_{c}$, albeit for a limited region of the spectrum. Level repulsion is one of the indicators of chaotic motion in the classical limit. This chaotic classical motion above the QPT also occurs in the Dicke model but it is interesting to note that it is totally absent in the purely bosonic case (no impurity) due to the latter's integrability even though it also features a ground state bifurcation.

We also found that self-trapping can occur in this system when $W> 2 W_{c}$ due purely to the boson-impurity interaction. We argued that the ``impurity-induced'' self-trapping states occur within the loops (like in the purely bosonic case)  and have a life-time scaling exponentially with the number $N$ of bosons and should be long-lived for a moderate number of atoms.

At the beginning of this paper we mentioned that the boson-impurity
system can be regarded as a poor man's Dicke model. The same qualitative behaviour occurs in both systems,
although the Dicke model is more sensitive to its interaction
parameter than the boson-impurity system as seen in Eqs.\
(\ref{eq:groundBI}) and (\ref{eq:groundDicke}).  Also, comparing
our level-spacing distributions to others obtained for the Dicke
model we see that there is a clearer change from Poissonian
to Wigner-Dyson in the Dicke model \cite{emary03,graham87}.  Nevertheless, it is remarkable that such a drastic reduction in the size of the Hilbert space to the absolute minimum preserves the critical features of the Dicke model. The lesson of this work is that all that is needed is the barest hint of an extra degree of freedom to simulate the presence of the harmonic oscillator. 

Finally, we note that the emergence of chaos heralds a second order QPT in the Dicke model, so the obvious question to ask is whether there is also a true QPT in boson-impurity model in the thermodynamic limit and whether it is in the same universality class as that of the Dicke model? Studies of the purely bosonic case suggest that this is likely \cite{botet83,shchesnovich09,oles10,buonsante11,juliadiaz10}. A modified version of the Dicke phase transition has recently been seen using cold atoms inside an optical cavity which is illuminated from the side by a laser \cite{baumann10}. Below the transition most of the light passes through the cavity but above it the atoms spontaneously form a matter-wave grating which efficiently scatters light into the cavity. The phase transition can be continuously observed by detecting the photons leaking through the end mirrors, but by the same token this means that the system is open and this modifies the critical behavior slightly \cite{nagy10,buchhold12}. By contrast, our system is closed (except for the insignificant rate of atom loss) and so it may in fact give a better match to the quantum properties of the Dicke model.

\begin{acknowledgments}
We thank Maxim Olshanii and Han Pu for useful discussions. JL acknowledges support from the Swedish research 
council (VR) and DO'D acknowledges support from NSERC (Canada).
\end{acknowledgments}

\appendix

\section{A basis with well-defined parity in Fock space}
\label{sec:appendixparity}

Numerical diagonalization routines do not generally respect the parity of
eigenstates that are nearly degenerate.  This directly impacts the
calculations of $\langle \hat{S}_z \rangle$ and the statistics of the
level spacings since we separate the hamiltonian into even and odd
parity blocks.  To overcome this obstacle we force the eigenstates to
have good parity (GP) by diagonalizing the hamiltonian in a basis with well-defined parity.  Instead of using the ``bare'' Fock basis $\vert \Delta N, \Delta M \rangle$, we use a basis whose states
are linear combinations of Fock states

\begin{equation}
\vert \mathrm{GP} \rangle = 
\begin{cases}
(\vert \Delta N, \Delta M \rangle + \vert - \Delta N,
- \Delta M \rangle)/\sqrt{2} \\
(\vert \Delta N, \Delta M \rangle - \vert - \Delta N,
- \Delta M \rangle)/\sqrt{2} 
\end{cases} \, .
\end{equation}
\\
After the diagonalization is complete we still want to represent the
eigenstates of the hamiltonian in the Fock basis, so we rotate the
parity states back with a unitary transformation

\begin{equation}
U^{\dagger} = \sum_{n=1}^{2N+2} \vert \mathrm{Fock}^{(n)} \rangle
\langle \mathrm{GP}^{(n)} \vert \, .
\end{equation}
\\
The Fock states now have good parity and can be used in our
calculations.

\begin{figure*}[t]
\begin{center}
\begin{tabular}{c}
\includegraphics[width=2.0\columnwidth]{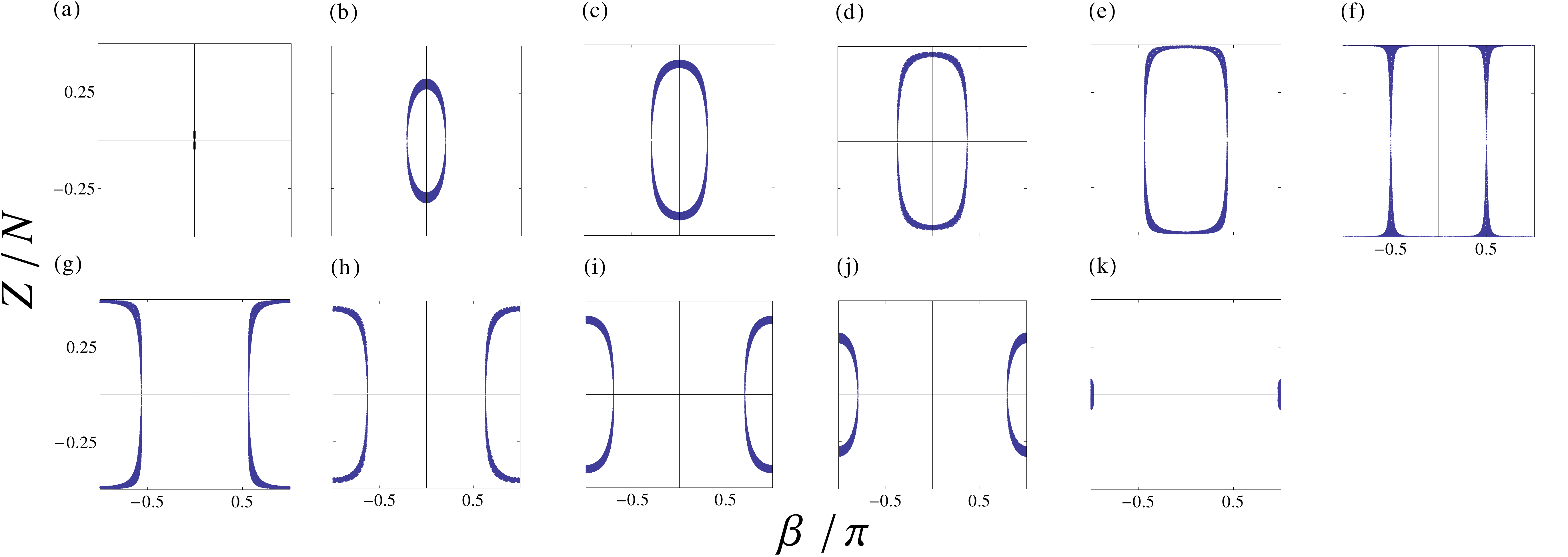} \\
\includegraphics[width=2.0\columnwidth]{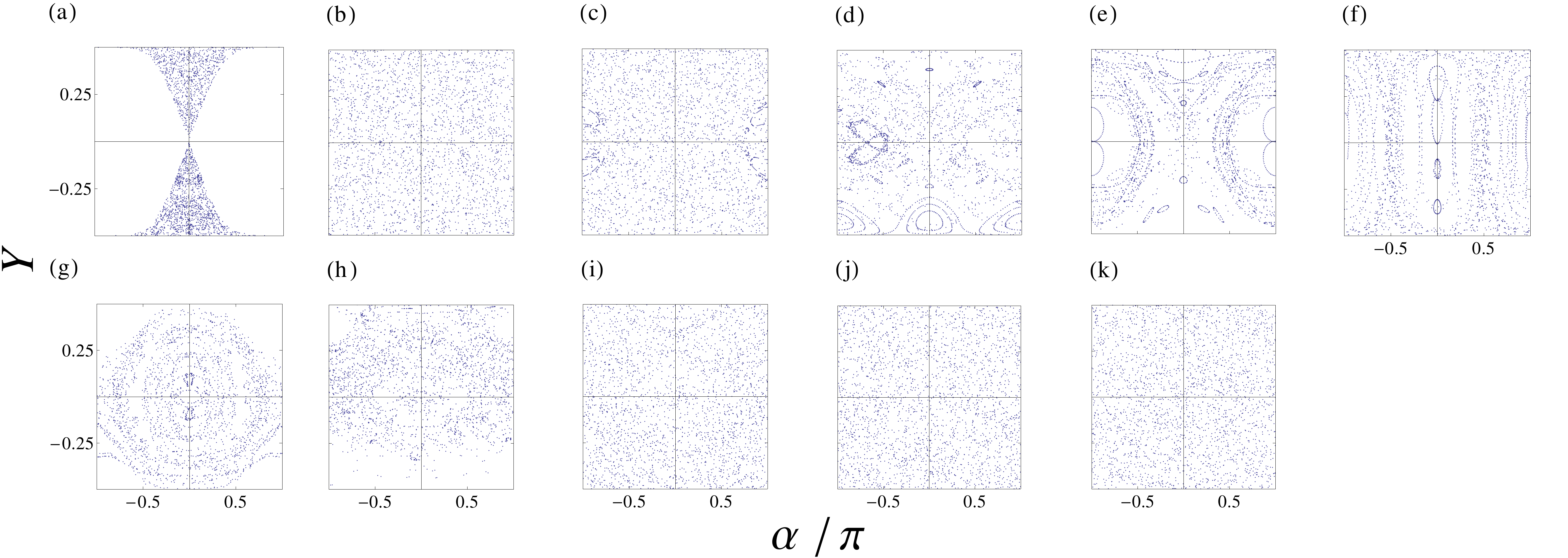}
\end{tabular}
\end{center}
	\caption[Poincar\'{e} sections for fixed $W$]{(Color online) Poincar\'{e} sections created by
          plotting intersections through a 2D plane of the 4D phase space.  For each plot,
          thirty random on-shell initial conditions are used and are
          evolved over a period $\tau=150$ ($\tau=Jt/\hbar$).  Each
          row shows intersections in different 2D planes: the top row is
          the $\frac{Z}{N}\frac{\beta}{\pi}$ plane with each point corresponding to $\alpha
          = 0$ and the bottom row is the $Y\frac{\alpha}{\pi}$ plane with each
          point corresponding to $\beta = 0$.  From left to right the
          plots increase in energy shell: (a)
          $E_{\mathrm{shell}} = -500J$, (b) $E_{\mathrm{shell}} =
          -400J$,
          (c) $E_{\mathrm{shell}} = -300J$, (d) $E_{\mathrm{shell}} = -200J$, (e) $E_{\mathrm{shell}} = -100J$, (f)
          $E_{\mathrm{shell}} = 0J$, (g) $E_{\mathrm{shell}} = 100J$, (h) $E_{\mathrm{shell}} = 200J$, (i) $E_{\mathrm{shell}} = 300J$, (j)
          $E_{\mathrm{shell}} = 400J$, and (k) $E_{\mathrm{shell}}
          = 500J$.  The values of the other
          parameters are $N=498$, $\Delta \epsilon=\Delta
          \epsilon^{a}=0$, $J^a = 2 J$ and $W = 1.5W_c$.}
	\label{fig:PPanel22}

\end{figure*}

\section{Poincar\'{e} sections for increasing energy shell}
\label{sec:app}
In this appendix we show Poincar\'{e} plots of dynamics on increasing
energy shells for the impurity and bosons.  The energy shell range is
$-JN - J^a \leq E_{\mathrm{shell}} \leq JN + J^a$ which covers the region between
the two loops as shown in plot (c) of Fig.\ \ref{fig:EvTilt}.
Figure \ref{fig:PPanel22} shows Poincar\'{e} plots for the bosons and
impurity.  Going from left to right each plot shows an
increase in the energy shell by $100J$ which is one tenth of the
range for $J^a = 2J$ and $N = 498$.  The value of $W$ is held
constant at $W = 1.5W_c$, so we are always in the chaotic regime. 
 
The region of phase space accessible to the bosons is restricted
whereas the impurity can access its entire phase
space.  This is due to the
small impurity hopping energy, $J^a$, relative to $JN$.  The bosons
can be thought of as a reservoir of energy, whose dynamic behaviour on a global
scale is barely affected by the impurity.  However, locally the
dynamics take place in a band whose thickness depends on $J^a$;
within the band there is chaos.

In the bottom two rows of Fig.\ \ref{fig:PPanel22} we see that the dynamics
of the impurity get less chaotic, up to a
point, as the energy shell is increased, then become more chaotic
again. To explain this we note that the top loop in Fig.\ \ref{fig:EvTilt} forms from a
supercritical bifurcation when $W >W_c$ (the top loop would form
sooner for $U \neq 0$), so both loops are symmetric in that regard.
This is why we get a symmetric degree of chaos around the plot that
corresponds to $E_{\mathrm{shell}} = 0$ (bottom plot (f)).  The
location in phase space of the second unstable point is
$(0,\pi,0,\pi)$.  Top plot (f) shows the dynamics of
the bosons when $E_{\mathrm{shell}} = 0$.  We see that when $\beta = 0$ or
$\beta = \pm \pi$, $Z = \pm N/2$ and when $Z = 0$, $\beta = \pm \pi/2$, so
the bosons are always at a maximal distance from the unstable points
in phase space.  When we increase $E_{\mathrm{shell}}$ further we see the
dynamics converge to regions around $(\beta,Z) = (\pm \pi,0)$ which causes
an increase in chaotic behaviour of the impurity.  The important point
to make here is that for our parameters the energy needed by the impurity to
access all of its phase space is small compared to the energy of the
entire system.  This means that the impurity has access to the
regions around $(\alpha, Y) = \{(0,0),(\pm \pi,0) \}$.  Therefore, the
degree of chaos of the impurity dynamics depends on how close the
bosons are to $ (\beta, Z) = \{(0,0),(\pm \pi,0) \}$


\begin{thebibliography}{8}
\bibitem{com}{Starting with all particles in the same well, impurity plus bosons, does not change the conclusions of the impurity induced self-trapping.}
\end{thebibliography}
\end{document}